\documentclass{aa}  
\usepackage{xcolor}
\usepackage{graphicx}
\usepackage{txfonts}
\usepackage{lipsum}
\usepackage{subcaption}       
\usepackage{lscape}             
\usepackage{placeins}          
\newcommand{\jl}[1]{#1}

\newif\ifshowchanges
\showchangesfalse
\ifshowchanges
    \newcommand{\rev}[1]{\textcolor{blue}{#1}}
\else
    \newcommand{\rev}[1]{#1}
\fi
                
\begin{document}
    \title{The Galactic disk structure revealed by Gaia XP spectra}

\author
{Zhenkang Tang\inst{1}\email{tangzhenkang@stu.ynu.edu.cn}
\and
Jianhui Lian\inst{1,2}\corrauth{jianhui.lian@ynu.edu.cn}
\and
Jialong Zhou\inst{1}\email{zhoujl@stu.ynu.edu.cn}
\and
Jiadong Li\inst{3}\email{jdli@mpia.de}
}

\institute{
South-Western Institute for Astronomy Research, Yunnan University,
Kunming, Yunnan 650091, People's Republic of China
\and
Yunnan Key Laboratory of Survey Science, Yunnan University,
Kunming, Yunnan 650500, People's Republic of China
\and
Max-Planck-Institut für Astronomie, Königstuhl 17, D-69117 Heidelberg, Germany
}
   \date{Received 16 July 2026 / Accepted 8 October 2026}

  \abstract
    {The spatial structures of different chemical populations in the Galactic disk provide important constraints on disk evolution and underlying physical processes. Gaia XP spectra provide one of the largest spectroscopic samples currently available, offering a new opportunity to investigate Galactic disk structure.}
    {We use Gaia XP spectra to study the structure of mono-abundance populations (MAPs) of the Galactic disk.
    In particular, we measure their scale heights as a function of radius as well as radial surface-mass-density profiles.}
    {We construct a red giant branch (RGB) sample from Gaia XP spectra and divide it into MAPs according to $[\alpha/{\rm M}]$ and $[{\rm M/H}]$. The working sample covers a $6\,{\rm kpc}\times6\,{\rm kpc}$ region centered on the Sun in the Galactic plane and extends to $|Z|<3.7\,{\rm kpc}$ in the vertical direction. We estimate the XP completeness using a Gaia photometric parent sample. We then combine PARSEC isochrones and a Kroupa initial mass function (IMF) to calculate the selection function and convert the observed stellar counts into intrinsic stellar mass densities. The vertical density profiles are fitted with an exponential model to derive the scale heights, and the corresponding surface mass densities are obtained by integrating the fitted vertical profiles.}
    {The scale height and surface-mass-density profiles derived from Gaia XP are generally consistent with those obtained from APOGEE. All MAPs show flaring, with the high-$\alpha$ population exhibiting the strongest flaring and the metal-poor and sub-solar low-$\alpha$ populations showing the weakest. The surface-mass-density profiles also differ significantly among MAPs. The high-$\alpha$ and metal-rich low-$\alpha$ populations show approximately exponential declines, whereas the other low-$\alpha$ populations exhibit plateau-like features, indicating that the Galactic disk cannot be described by a simple single-exponential profile.}
    {The differential structures of MAPs suggest a complex assembly history of the Galactic disk likely regulated by birth conditions, secular dynamical heating and radial migration, and external perturbations.}

   \maketitle
   \nolinenumbers

\section{Introduction}

    The spatial structure of the Milky Way disk contains important information about the formation and evolution of the Galaxy. Traditionally, the Galactic disk has been decomposed into two main components, the thin disk and the thick disk. These two components differ significantly in scale height, age, metallicity, $\alpha$-element abundance, and kinematic properties: the thick disk is generally older, vertically thicker, more metal-poor, more $\alpha$-enhanced, and dynamically hotter, while the thin disk is typically younger, thinner, and dominated by low-$\alpha$ populations \citep{Yoshii1982,Gilmore1983,Fuhrmann1998,Robin2003,Bensby2005,Juric2008,BlandHawthorn2016}. \rev{With the advent of large-scale spectroscopic surveys such as the Apache Point Observatory Galactic Evolution Experiment (APOGEE), the Large Sky Area Multi-Object Fiber Spectroscopic Telescope (LAMOST), the Galactic Archaeology with HERMES (GALAH) Survey, and Gaia}, it has become clear that Galactic disk stars show a prominent chemical bimodality in the $[\alpha/{\rm Fe}]$--$[{\rm Fe/H}]$ abundance plane, namely the high-$\alpha$ and low-$\alpha$ sequences \citep{Fuhrmann1998,Reddy2006,Lee2011,Adibekyan2012,Haywood2013,Bensby2014,Nidever2014,Hayden2015,Queiroz2020}. This chemical bimodality not only reflects the chemical-evolution history of the Galactic disk, but is also closely related to the ages, spatial distributions, and kinematic properties of different stellar populations \citep{Imig2023}. Therefore, the vertical structure, including the scale height and its spatial variation, and the radial structure, traced by surface-density profiles, provide critical observational constraints on the formation history of the Galactic disk and the underlying astrophysical drivers \citep{Robin2003,Juric2008}.

    Compared with the traditional geometric decomposition into thin and thick disks, mono-abundance populations (MAPs), defined in the $[\alpha/{\rm Fe}]$--$[{\rm Fe/H}]$ plane, provide a more physically informative way to divide disk populations and to trace the evolutionary history of the Galaxy \citep{Bovy2012a,Bovy2016,Minchev2016,Mackereth2017,Lian2022a}. Early studies of MAPs showed that the Galactic disk cannot be described simply by two geometric components, but is instead composed of a series of chemical populations with continuously varying structural properties \citep{Bovy2012a,Bovy2016}. \rev{\citet{Bovy2012a} modeled the spatial structure of MAPs using G dwarfs from the Sloan Extension for Galactic Understanding and Exploration (SEGUE), a component of the Sloan Digital Sky Survey (SDSS)} and found that the scale heights and radial scale lengths of different MAPs vary systematically with chemical abundance. Old, $\alpha$-enhanced, and metal-poor populations generally have larger scale heights and shorter radial scale lengths, while younger low-$\alpha$ populations are more radially extended \citep{Mackereth2017}. Similarly, studies of mono-age populations based on LAMOST have further shown that the radial and vertical structures of the Galactic disk vary significantly with stellar age \citep{Xiang2018,Yu2025,Xiang2025}. \citet{Xiang2018} reconstructed the three-dimensional mass distributions of mono-age populations using LAMOST \rev{main-sequence turnoff and subgiant branch stars} and found that populations younger than $10\,{\rm Gyr}$ commonly show flaring, that is, an increase in disk scale height with Galactocentric radius. They also found that the radial scale length of the disk varies with age, and that these trends support an inside-out formation picture for the Galactic disk. More recently, \citet{Xiang2025} used high-$\alpha$ stars with significant \rev{prograde} orbital angular momentum \rev{($L_Z>500\,{\rm kpc.km/s^{-1}}$) to select a rotating, disk-like population with precise age estimates and investigate the age-dependent structure of the old high-$\alpha$ disk}. They found that its radial scale length is nearly independent of age, whereas its vertical scale height evolves strongly with age, and that a disk-like geometry is already present for stellar populations older than $13\,{\rm Gyr}$. This result suggests that the Milky Way formed a surviving old stellar disk at a very early epoch. APOGEE data have also been used to reconstruct the intrinsic density distributions of MAPs after carefully accounting for the survey selection function. \citet{Lian2022a} confirmed that low-$\alpha$ populations show broken radial profiles and flaring, consistent with earlier APOGEE studies \citep{Bovy2016,Mackereth2017}. However, they found that the high-$\alpha$ MAPs exhibit the strongest flaring, in contrast to the nearly constant thickness of high-$\alpha$ populations inferred by \citet{Bovy2016}.
    
   Recent studies have revealed that the Milky Way disk contains structural components that are not easily described by the traditional thin/thick  disk picture. \rev{For example, \citet{Binney2024} constructed a chemodynamical model with multiple stellar disk components using APOGEE and Gaia data, while \citet{Khanna2025} fitted a two-component disk model to Gaia red-clump stars, finding different structural and flaring properties between the two components.} Based on APOGEE and Gaia data, \citet{Lian2025} identified a relatively young thick-disk component with an age of about $6.6\,{\rm Gyr}$. This component belongs to the metal-poor low-$\alpha$ population and has a scale height and flaring strength comparable to those of the canonical high-$\alpha$ thick disk, while being more radially extended and systematically younger. This finding suggests that secular dynamical heating alone may not explain the formation of thick-disk structures. Instead, both the canonical old high-$\alpha$ thick disk and the younger metal-poor low-$\alpha$ thick disk may have formed through similar turbulent and bursty birth conditions, possibly triggered by galaxy interactions or merger events.
    
    Recent measurements have also challenged the traditional view that the radial structure of the Milky Way disk can be described by a single-exponential profile \citep{Sackett1997,LopezCorredoira2002,Reyle2009,BlandHawthorn2016}. Based on APOGEE data, \citet{Lian2024} constructed an age-resolved luminosity surface-density profile and found that the Milky Way disk exhibits a nearly flat plateau over $3.5<R<7.5\,{\rm kpc}$, followed by a clearer exponential decline at larger radii. More recently, \citet{Wu2026} used the Gaia BP/RP red-clump sample to analyze the Galactic disk density profile and confirmed the presence of a similar inner plateau. These results imply that the global radial structure and size of the Milky Way may be more complex than expected from a single-exponential disk model. For example, \citet{Lian2024} suggested a significantly larger Galactic disk, while \citet{Lian2025b} found that the inferred stellar mass is about a factor of two lower than previous estimates. Therefore, an independent test of the radial mass distributions and vertical structures of different MAPs using a large stellar sample is important for understanding the intrinsic structure and formation history of the Galactic disk.

    The Gaia XP spectroscopic sample provides a new opportunity for this purpose. \rev{The third data release of the Gaia (Gaia DR3)} provides large-scale low-resolution XP spectra, making it possible to study the chemical structure of the Galactic disk with a much larger sample and more continuous spatial coverage \citep{Gaia_Collaboration2023,DeAngeli2023}. In addition, unlike many ground-based spectroscopic surveys, Gaia XP is based on Gaia's all-sky scanning observations and does not rely on complex target-selection strategies. Gaia XP has therefore become one of the most important data sets for separating different chemical populations over a wide spatial range. However, the use of Gaia XP data for Galactic structure studies also faces several key challenges. First, the XP spectroscopic sample is not fully complete, and its selection is affected by magnitude, sky position, the number of Gaia transits, and spectral quality \citep{Andrae2023,DeAngeli2023}. Existing Gaia XP studies have provided valuable constraints on Galactic structure and have improved stellar-parameter estimation, chemical-abundance calibration, and empirical completeness corrections \citep{Widmark2022,Wu2025}. For reconstructing the intrinsic three-dimensional mass-density distributions of different MAPs, however, one must further account for selection effects introduced by the sample selection criteria and different stellar properties (i.e., age and metallicity). Because these properties affect the fraction of the underlying population represented by the observed tracers, structures inferred from stellar distributions corrected only for incompleteness may still be biased. We therefore combine the Gaia XP completeness with a selection function based on stellar evolutionary models and an initial mass function to convert observed stellar counts into intrinsic stellar mass densities \citep{Lian2022a}. Such intrinsic density profiles further provide a way to connect Galactic archaeology with the integrated properties and mass assembly history of the Milky Way as a galaxy \citep{Zasowski2025,Lian2024}.
    
    Our work aims to test the reliability of Gaia XP for Galactic disk structure studies and to further understand the radial and vertical structural evolution of the Galactic disk within a few kpc to the Sun. In this way, we use the large Gaia XP sample to study the three-dimensional mass density structures of different MAPs. We combine stellar evolutionary models to calculate the selection function in each spatial element, and convert the observed stellar counts into stellar mass densities. Based on these reconstructed density distributions, we measure the vertical scale heights and surface mass densities of the high-$\alpha$ population and different low-$\alpha$ MAPs as functions of Galactocentric radius, with a particular focus on the flaring and radial mass distributions of different chemical populations. Finally, we compare these Gaia XP-based measurements with previous APOGEE results to assess the robustness of the reconstructed structures.

\section{Data}

    \subsection{Gaia XP spectra}
    \label{sec:xp_spectra}
    In this work, we use the Gaia BP/RP spectroscopic sample from the Gaia DR3 mission \citep{Gaia_Collaboration2016,Gaia_Collaboration2023} as the primary dataset to investigate the spatial structure of the Galactic disk. The Gaia DR3 provides a large number of low-resolution BP/RP (XP) spectra, usually referred to as Gaia XP spectra. These spectra are obtained with the BP and RP photometer, covering a wavelength range of approximately $330{-}1050\,\mathrm{nm}$ with a resolving power of $R\simeq15{-}85$ \citep{Andrae2023}. Gaia XP spectra offer two major advantages. First, their sample size far exceeds that of existing spectroscopic surveys. Gaia DR3 provides XP spectra for approximately 220 million stars, making it one of the largest homogeneous spectroscopic datasets currently available. Second, thanks to the precise photometry and astrometric measurements delivered by Gaia, XP spectra can be naturally combined with distance, kinematic, and orbital information, providing a unified data set for Galactic studies. Consequently, Gaia XP spectra have been widely used to investigate the structure and chemical evolution of the Galactic disk \rev{\citep{Kurbatov2024,Chandra2024,Andrae2023b}}.

    In this work, we adopt the Gaia XP stellar labels derived by \citet{Li2024}. \citet{Li2024} used the AspGap framework, a data-driven method based on neural networks, which was trained on stars in common between Gaia XP and APOGEE \rev{Data Release 17}. This framework enables the direct inference of stellar parameters and chemical abundances from the low-resolution Gaia XP spectra, including the effective temperature $T_{\rm eff}$, surface gravity $\log g$, metallicity $[\mathrm{M/H}]$, and $\alpha$-element abundance $[\alpha/\mathrm{M}]$. 
    
    Although XP spectra have a very low resolving power, they still carry substantial chemical information: strong, broadened, and blended absorption features remain imprinted over many pixels, and Gaia's accurate absolute flux calibration preserves the overall spectral energy distribution (SED) shape, which is sensitive to metallicity through metal line blanketing \citep{Zhang2023}. Using this information, \citet{Zhang2023} derived stellar parameters and metallicities for $\sim$220 million stars from XP spectra, while \citet{Laroche2025}, adopting an unsupervised and stellar-label-independent approach, further demonstrated that XP spectra also encode $\alpha$-element abundance information by recovering the $\alpha$-bimodality directly from the spectra. These results motivate our use of the XP-based stellar labels of \citet{Li2024} to separate chemically distinct populations of the Galactic disk.
    
    \citet{Li2024} applied a series of quality-control and evolutionary-stage cuts to build their published Red Giant Branch (RGB) catalog. While these cuts improve sample purity for abundance studies, they are unsuitable here: they introduce non-negligible selection effects, and the RGB selection is partly based on $T_{\rm eff}$, which is unavailable for our Gaia photometric parent sample (Sect.~\ref{sec:sample_selection}) and thus cannot be applied consistently to both samples. We therefore instead use the full AspGap label catalog of \citet{Li2024}, which omits these final cuts and covers nearly all Gaia XP sources, and apply our own RGB selection identically to the XP and photometric samples, enabling a self-consistent completeness correction. Because our spatial and RGB selections rely on stellar colors and absolute magnitudes, which are affected by interstellar extinction, we first correct the photometry for extinction and perform all subsequent selections on the extinction-corrected quantities.

    \begin{figure*}[ht!]
    \sidecaption
    \includegraphics[width=12cm]{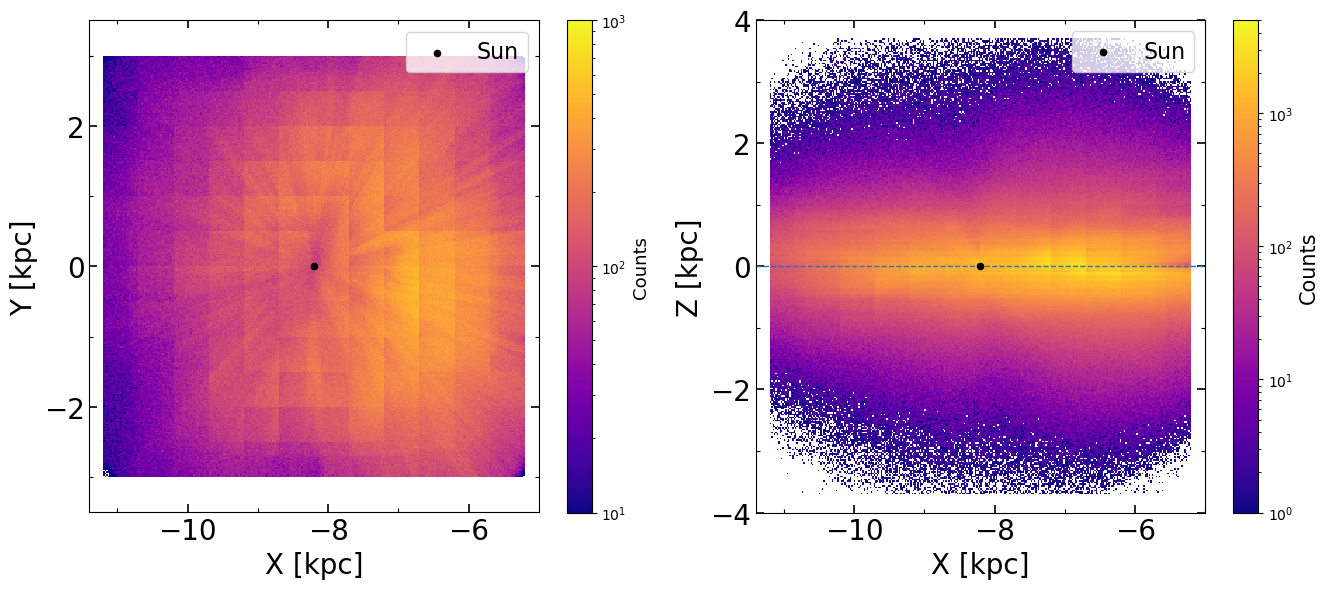}
    \caption{
    Distribution of the extinction-corrected sample in the Galactocentric coordinate system. Left: stellar number density distribution in the Galactic $XY$ plane. The sample covers a $6\times6 \,\mathrm{kpc}^2$ region centered on the Sun. Right: stellar number density distribution in the Galactic $XZ$ plane, spanning $|Z|<3.7\,\mathrm{kpc}$. Only spatial elements whose eight vertices lie entirely within the adopted $4.5\,{\rm kpc}$ heliocentric sphere are retained in the analysis.
    }
    \label{figdis}
    \end{figure*}

    \subsection{Sample selection and extinction correction}
    \label{sec:sample_selection}
    In this work, stellar distances are estimated from Gaia parallaxes after applying the Gaia DR3 parallax zero-point correction. We first restrict the sample to stars with relative parallax uncertainties smaller than 20\% in order to ensure reliable distance estimates. The parallaxes are then corrected for the global Gaia DR3 parallax zero-point following the prescription of \citet{Lindegren2021}, and distances are calculated by directly inverting the corrected parallaxes. We further restrict the sample to stars located within a sphere of radius $4.5\,{\rm kpc}$ centered on the Sun. For the extinction correction, completeness correction, and subsequent density reconstruction, the analysis is performed independently in a series of 3D spatial elements based on the same binning scheme. In the Galactocentric coordinate system, we consider a $6\,{\rm kpc}\times6\,{\rm kpc}$ region centered on the Sun in the $XY$ plane and divide it into spatial cells of size $0.5\,{\rm kpc}\times0.5\,{\rm kpc}$. The Sun is located at $(X,Y,Z)=(-8.2,0,0.027)\,{\rm kpc}$ and is placed at the corner shared by four adjacent $XY$ cells \citep{Chen2001,Gravity2019}. In the vertical direction, the sample is further divided into height bins according to $|Z|$, with bin boundaries of $0$, $0.1$, $0.2$, $0.3$, $0.4$, $0.6$, $0.8$, $1.0$, $1.4$, $1.8$, $2.2$, $2.7$, $3.2$, and $3.7\,{\rm kpc}$. The same binning scheme is applied above and below the Galactic mid-plane. We only retain spatial elements whose eight vertices all lie within the adopted $4.5\,{\rm kpc}$ sphere. After this requirement, the final spatial grid contains 3656 spatial elements.
    
    We adopt the all-sky three-dimensional extinction map constructed by \citet{Guo2021} for extinction correction. This map combines a southern-sky three-dimensional extinction map derived from multiple photometric surveys with previously published extinction maps for the northern sky and the Galactic plane, providing nearly full-sky coverage. The extinction map is based on stellar spectral energy distribution fitting and incorporates Gaia distance information to reconstruct the three-dimensional distribution of dust along different lines of sight. To convert the extinction-map measurements into the Gaia photometric system, \rev{we adopt the extinction coefficients based on the Model Atmospheres in Radiative and Convective Scheme stellar atmosphere models of \citet{Casagrande2018}. Since the extinction coefficients depend on stellar parameters and reddening, we derive sample-specific mean coefficients using their interpolation routines. We divide our XP sample into a series of bins in $T_{\rm eff}$, $\log g$, $[\mathrm{M/H}]$, and $E(B-V)$ over the parameter ranges covered by our sample, calculate the extinction coefficients for each bin, and derive the mean coefficients by weighting each bin according to the number of stars it contains. This yields $R_G=2.516$, $R_{BP}=3.173$, and $R_{RP}=1.974$.}
    
    The extinction correction is performed on the same spatial grid described above. Within each $XY$ cell, 25 positions are uniformly sampled in the $XY$ plane. Along the vertical direction, we use the midpoint of each height bin as the representative sampling height. This gives a total of 91\,400 sampling points. For each spatial element, the extinction values are obtained by querying the extinction map at the corresponding sampling points, and the mean value is adopted as the representative extinction of that spatial element. All stars located in the same spatial element are then assigned the same extinction correction. This approach reduces the computational cost of star-by-star extinction queries while keeping the extinction correction consistent with the spatial resolution used for the completeness and density calculations.

    \begin{figure*}[ht!]
    \sidecaption
    \includegraphics[width=12cm]{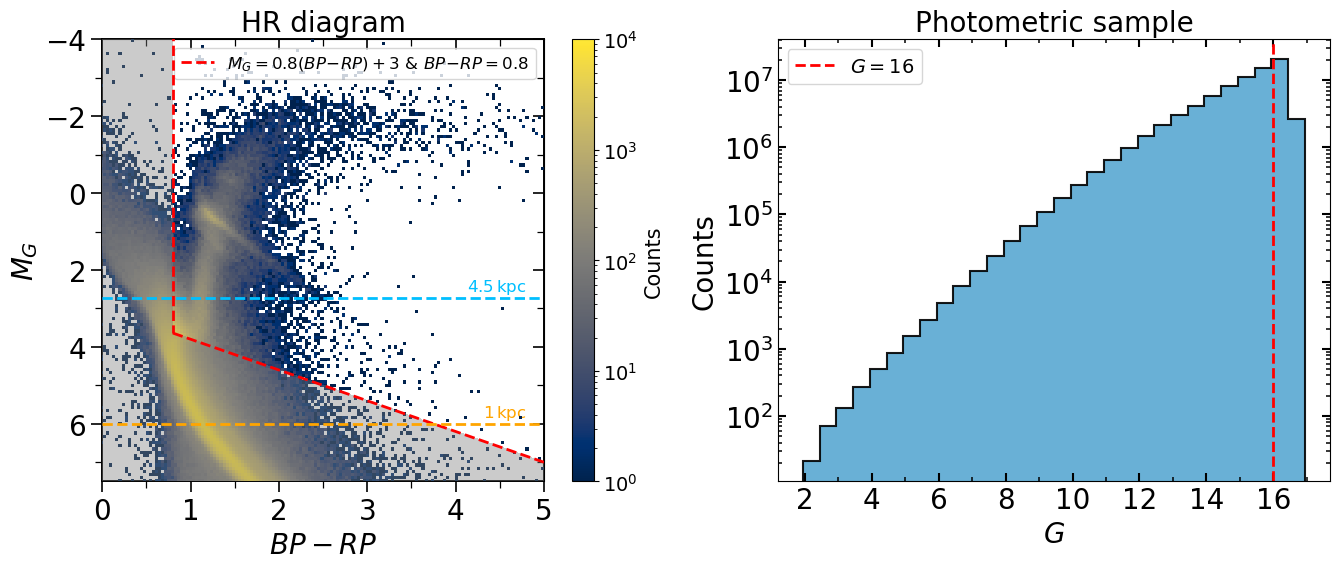}
    \caption{
    Illustration of the adopted RGB selection criterion and the magnitude distribution of the parent photometric sample. Left: Hertzsprung–Russell diagram constructed from stars within a local cuboid defined by $-8.2\le X<-7.7 \,\mathrm{kpc}$, $0\le Y<0.5\,\mathrm{kpc}$, and $|Z|<3.7\,\mathrm{kpc}$. The red dashed lines indicate the color--magnitude selection criterion adopted for a spatial element with zero extinction ($A_G=0$). This criterion defines the final RGB sample used in the zero-extinction region. The light-blue horizontal line marks the absolute magnitude corresponding to the apparent magnitude limit of $G=16$ at a heliocentric distance of $4.5\,{\rm kpc}$, while the orange horizontal line indicates the corresponding limit at a distance of $1\,{\rm kpc}$. Right: apparent-magnitude distribution of the parent photometric sample within the adopted spatial volume. The red dashed line marks the magnitude limit $G=16$, adopted to ensure a high level of sample completeness.
    }
    \label{fig1}
    \end{figure*}
    
    To maintain a high level of photometric completeness within the adopted spatial volume, we restrict the Gaia XP sample to $6<G\leq16$. The bright limit corresponds to the brightest magnitude knot of the parallax zero-point model of \citet{Lindegren2021}, while the faint limit is adopted to ensure high photometric completeness. As discussed in Sect.~\ref{sec:xp_spectra}, stellar parameters and chemical abundances inferred from Gaia XP spectra are less reliable for dwarf stars \citep{Holtzman2018,Andrae2023b,Yao2024}. Therefore, after applying the extinction correction described above, we select RGB stars using a criterion in the extinction-corrected absolute magnitude–color plane. As shown in the left panel of Fig.~\ref{fig1}, we first define a reference RGB selection in a zero-extinction region as
    \begin{equation}
    \label{eqmg}
    M_G \leq 0.8(BP-RP)+3 ,
    \end{equation}
    together with
    \begin{equation}
    \label{eqc}
    BP-RP>0.8 .
    \end{equation}
    This criterion corresponds to the region above and to the right of the red dashed line, namely the area not covered by the gray mask. \rev{Before constructing the final sample, we also cross-match the RGB stars with the globular-cluster membership catalogs of \citet{Vasiliev2021} and remove 3616 stars with membership probabilities of $P_{\rm mem}\geq90\%$ to minimize contamination from known Galactic globular clusters. Applying all of the above criteria yields a final RGB sample of 12,170,924 stars. For the subsequent analysis, we retain only spatial elements whose eight vertices lie entirely within the adopted $4.5\,{\rm kpc}$ heliocentric sphere.} The spatial distributions of this sample in the Galactocentric $XY$ and $XZ$ planes are shown in the Fig.~\ref{figdis}.

    \subsection{Mono-abundance populations}
    Stellar chemical abundances provide important information on stellar birth environments and Galactic evolution. To investigate the structural properties of stellar populations with different chemical abundances, we follow previous studies and divide the sample into MAPs in the chemical-abundance plane.

    Specifically, the sample is first separated into high-$\alpha$ and low-$\alpha$ populations in the $[\alpha/{\rm M}]$–$[{\rm M/H}]$ plane. Following the chemical-sequence division of \citet{Adibekyan2011}, we adopt a recalibrated boundary optimized for the abundance distribution of our Gaia XP sample. To minimize contamination caused by abundance uncertainties between the two chemical sequences, we do not define the populations using a continuous set of MAPs. Instead, a gap is introduced around the dividing boundary, and only stars lying sufficiently above or below the boundary are assigned to the high-$\alpha$ and low-$\alpha$ populations, respectively. This selection yields chemically cleaner samples at the expense of excluding a small fraction of stars near the transition region. The resulting boundary is defined as
    \begin{equation}
    [\alpha/{\rm M}]_{\rm high} =
    \begin{cases}
    0.24, & [{\rm M/H}] < -0.5,\\
    -0.240[{\rm M/H}] + 0.12, & -0.5 \leq [{\rm M/H}] < 0,\\
    \end{cases}
    \label{eq:alpha_high}
    \end{equation}
    and the low-$\alpha$ boundary is defined as
    \begin{equation}
    [\alpha/{\rm M}]_{\rm low} =
    \begin{cases}
    0.12, & [{\rm M/H}] < -0.3,\\
    -0.222[{\rm M/H}] + 0.05, & -0.3 \leq [{\rm M/H}] < 0,\\
    0.05, & [{\rm M/H}] \geq 0 .
    \end{cases}
    \label{eq:alpha_low}
    \end{equation}
    
    Stars located above Eq.~(\ref{eq:alpha_high}) are classified as high-$\alpha$ stars, while stars below Eq.~(\ref{eq:alpha_low}) are classified as low-$\alpha$ stars. In addition, we require $[{\rm M/H}]<0$ for the high-$\alpha$ population and restrict the low-$\alpha$ population to the metallicity range $-0.7<[{\rm M/H}]<0.5$. Applying these criteria yields a final sample of \rev{2,095,627} high-$\alpha$ stars and \rev{6,735,570} low-$\alpha$ stars.
    
    \begin{figure*}[ht!]
    \sidecaption
    \includegraphics[width=12cm]{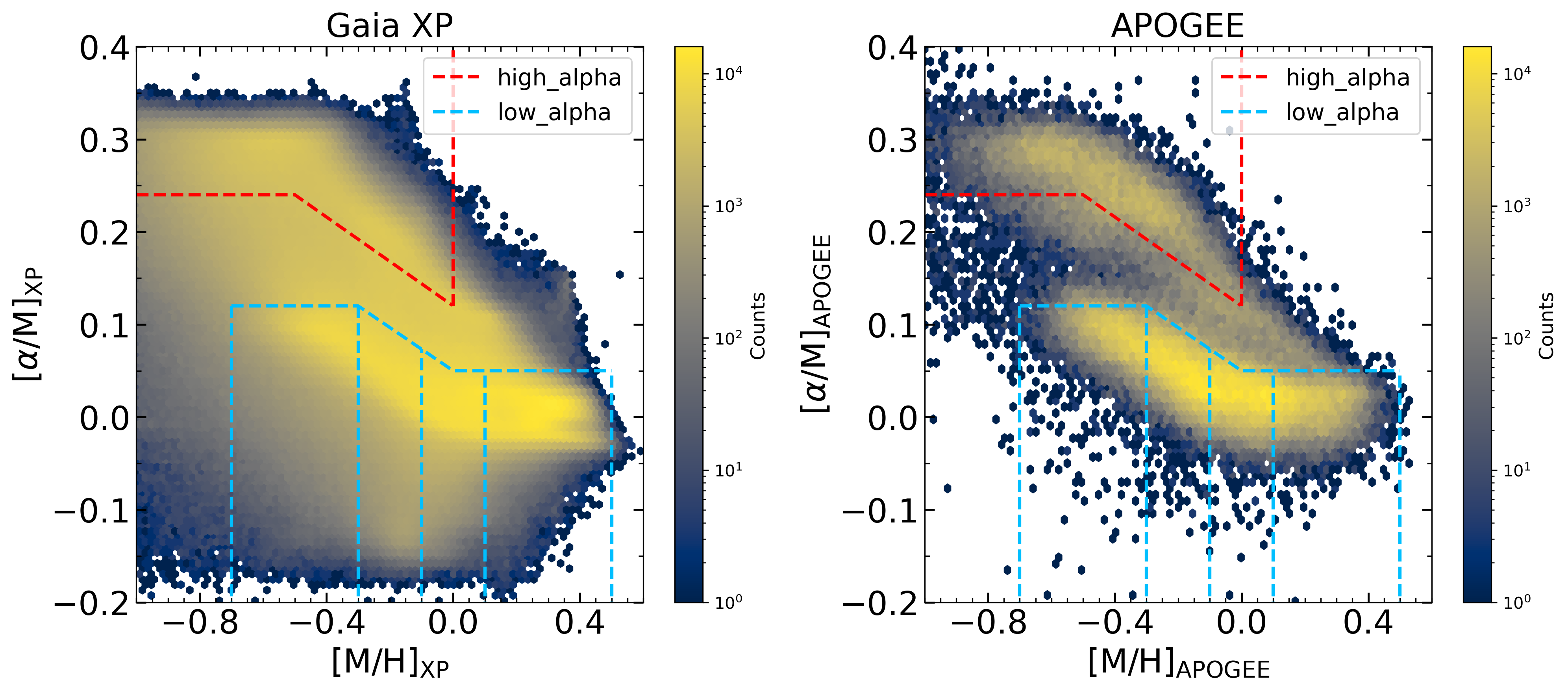}
    \caption{
    Chemical-abundance classification adopted in this work and comparison with the APOGEE abundance distribution. Left: distribution of the Gaia XP sample in the $[\alpha/{\rm M}]$–$[\mathrm{M}/\mathrm{H}]$ plane. The red and blue dashed lines indicate the selection criteria for the high-$\alpha$ and low-$\alpha$ populations, respectively. The blue rectangular cells define the MAPs used in the subsequent analysis. Right: distribution of the solar-neighborhood APOGEE sample in the $[\alpha/\rm{M}]$–$[\mathrm{M}/\mathrm{H}]$ plane. The same classification criteria are overplotted for comparison.
    }
    \label{fig2}
    \end{figure*}
    
    The low-$\alpha$ sequence is further divided into four metallicity intervals: $-0.7<[{\rm M/H}]<-0.3$, $-0.3<[{\rm M/H}]<-0.1$, $-0.1<[{\rm M/H}]<0.1$, and $0.1<[{\rm M/H}]<0.5$. Each interval is treated as an independent MAP. The adopted chemical-abundance classification and the resulting MAP definitions are illustrated in the left panel of Fig.~\ref{fig2}, where the red and blue dashed lines indicate the high-$\alpha$ and low-$\alpha$ boundaries, respectively, and the blue boxes mark the four low-$\alpha$ MAPs.
    
    For each MAP, we calculate the mean values of $[\alpha/{\rm M}]$ and $[{\rm M/H}]$ from all Gaia XP stars belonging to that population and adopt these mean abundances as the representative abundance values of the corresponding MAP for the subsequent analysis. The representative abundances of the five MAPs are listed in Table~\ref{tab:maps}.
    
    \begin{table}[ht!]
    \caption{\label{tab:maps}Definition and representative abundances of the MAPs used in this work.}
    \centering
    \begin{tabular}{lccc}
    \hline\hline
    Population & $[{\rm M/H}]$ range & $\langle[\alpha/{\rm M}]\rangle$ & $\langle[{\rm M/H}]\rangle$ \\
    \hline
    High-$\alpha$  & $[{\rm M/H}]<0.0$    & 0.20 & $-0.40$ \\
    Metal-poor low-$\alpha$ & $[-0.7,-0.3]$ & 0.10 & $-0.40$ \\
    Sub-solar low-$\alpha$ & $[-0.3,-0.1]$ & 0.00 & $-0.20$ \\
    Solar-metallicity low-$\alpha$ & $[-0.1,0.1]$ & 0.00 & 0.00 \\
    Metal-rich low-$\alpha$ & $[0.1,0.5]$ & 0.00 & 0.20 \\
    \hline
    \end{tabular}
    \end{table}

\section{Completeness and Selection function}

    \subsection{Completeness Correction}
    The availability and precision of Gaia XP labels depend on the quality of the underlying XP spectra. Since the XP spectral quality varies with the number of Gaia transits, sky position, and source brightness, the Gaia XP sample is affected by spatially varying selection effects and incompleteness \citep{DeAngeli2023, Andrae2023}. To correct for this incompleteness, we use the Gaia photometric sample as the parent sample and estimate the completeness of the XP sample relative to it. As shown in the right panel of Fig.~\ref{fig1}, we first examine the $G$-band magnitude distribution of the parent photometric sample within the adopted spatial volume. The number of stars increases continuously toward fainter magnitudes at $G\leq16$, with no clear truncation or turnover. We therefore regard the Gaia photometric sample as highly complete within this magnitude range \citep{Ivezic2019}. We also verify that the completeness of the photometric sample at $G\leq16$ is maintained at different spatial locations. Based on this consideration, we impose the same magnitude limit of $G\leq16$ on both the Gaia XP sample and the parent photometric sample.

    In principle, the completeness of the Gaia XP sample may depend on spatial position, absolute magnitude, and color. In our work, the XP completeness is estimated by comparing the number of Gaia XP stars with the number of stars in the parent photometric sample that satisfy the same selection criteria. The completeness can be expressed as a function of spatial position, absolute magnitude, and color:
    \begin{equation}
    f_{\rm comp}(X,Y,Z,\rev{M_G(X,Y,Z)},BP-RP)=
    \frac{N_{\rm XP}(X,Y,Z,M_G,BP-RP)}
    {N_{\rm phot}(X,Y,Z,M_G,BP-RP)},
    \label{eq:fcomp_general}
    \end{equation}
    where $N_{\rm XP}$ and $N_{\rm phot}$ denote the numbers of XP and photometric stars. To characterize the spatial dependence of the completeness, we evaluate $f_{\rm comp}$ within the three-dimensional spatial elements defined in Sect.~\ref{sec:sample_selection}. We then examine whether, within individual spatial elements, the completeness shows significant dependence on absolute magnitude $M_G$ and color $BP-RP$.

    Using the extinction corrections described above, we correct both the photometric and XP samples for extinction. In each spatial element, the same RGB selection criteria, given by Eqs.~(\ref{eqmg}) and~(\ref{eqc}), are then applied to both samples. We subsequently divide the absolute magnitude–\rev{color} plane into a set of small cells and calculate $N_{\rm XP}/N_{\rm phot}$ in each cell, where $N_{\rm XP}$ and $N_{\rm phot}$ are the numbers of XP and photometric stars in the corresponding cell. This quantity provides an estimate of the XP completeness as a function of absolute magnitude and color within each spatial element. Our tests show that the completeness distribution in the absolute-magnitude–color plane is relatively uniform for the selected RGB sample and does not exhibit significant systematic variations with absolute magnitude or color.
    
    Therefore, to improve the statistical robustness of the correction and to avoid introducing additional noise from sparsely populated cells, we only consider the spatial dependence of the XP completeness in the final correction. For each three-dimensional spatial element, we define the XP completeness as
    \begin{equation}
    f_{\rm comp}=\frac{N_{\rm XP,RGB}}{N_{\rm phot,RGB}},
    \label{eq:fcomp}
    \end{equation}
    where $N_{\rm XP,RGB}$ and $N_{\rm phot,RGB}$ are the numbers of XP and photometric stars satisfying the same RGB selection criteria within that spatial element \citep{Bovy2012b}. All XP stars located in the same spatial element are assigned the same completeness value $f_{\rm comp}$.

    \subsection{Selection Function and Stellar Mass Density Reconstruction}
    Because the final sample is selected using both the RGB criterion and the apparent magnitude limit of $6<G\leq16$, the observed RGB stars represent only a fraction of the underlying stellar population \citep{Lian2022a}. To reconstruct the intrinsic stellar mass density distributions of different MAPs, we therefore calculate a selection function that accounts for the combined effects of the RGB selection and the magnitude limit. The selection function is evaluated using the same three-dimensional spatial binning scheme described in Sect.~\ref{sec:sample_selection} and is calculated independently for each MAP within each spatial element.

    For the age distribution, we use the stellar ages and distances from the astroNN Value Added Catalog \jl{of APOGEE giant stars} to construct the age distributions of different MAPs as functions of Galactocentric radius and height above the Galactic plane \citep{Mackereth2019}\rev{. For most spatial bins, the adopted age profiles are directly constrained by stars with astroNN age measurements. For a small number of sparsely populated bins at large radii and high $|Z|$ in the metal-rich low-$\alpha$ population, the age profiles are estimated from neighboring populated spatial bins. We have verified that excluding these extrapolated age profiles does not affect our main results, except that a few outermost radial bins cannot be reliably fitted because of the reduced number of available spatial cells. For the metal-poor and sub-solar low-$\alpha$ MAPs, whose representative metallicities do not exactly coincide with the metallicity grid of the age profiles, we linearly interpolate between the two adjacent metallicity profiles.} Together with the representative $[\alpha/{\rm M}]$ and $[{\rm M/H}]$ values of the MAP, we then use PARSEC stellar evolutionary tracks \citep{Bressan2012,Marigo2013,Tang2014,Chen2014,Chen2015,Rosenfield2016,Pastorelli2019,Pastorelli2020} and a Kroupa Initial Mass Function (IMF, \citealt{Kroupa2001}) to model the distribution of stars in the color--magnitude diagram.
    
    Specifically, for a given MAP and spatial element, isochrones of different ages are weighted according to the assigned age distribution, and the stellar weights along each isochrone are computed using the IMF. We then apply the same RGB selection criteria as used for the observed sample, namely Eqs.~(\ref{eqmg}) and~(\ref{eqc}). For the apparent magnitude limit of $G\leq16$, the sample selection is defined in the observed-magnitude space, while the selection function is calculated from isochrones in the absolute-magnitude space. One possible approach would be to convert the isochrones into apparent magnitudes and apply the magnitude limit directly. However, the selection function is evaluated for finite spatial elements. At small distances, even modest distance variations can produce noticeable changes in apparent magnitude, introducing additional uncertainties into the selection-function calculation. We therefore first shift the isochrones according to the representative extinction of the corresponding spatial element and then convert the apparent magnitude limits into absolute-magnitude limits using the minimum and maximum heliocentric distances of that spatial element. The faint limit ($G=16$) is converted using the maximum heliocentric distance of the spatial element, while the bright limit ($G=6$) is converted using the minimum heliocentric distance. This guarantees that stars satisfying the adopted faint limit remain observable throughout the entire spatial element, while stars satisfying the adopted bright limit remain fainter than the bright-end cutoff even at the nearest location within the spatial element. Combining these limiting absolute magnitudes with the RGB selection criteria gives the total isochrone weight that satisfies the observational selection. The selection function $f_{\rm sel}$ is defined as the fraction of the IMF- and age-weighted stellar population that satisfies the adopted observational selection criteria for a given MAP and spatial element.

    \begin{figure*}[ht!]
    \sidecaption
    \includegraphics[width=12cm]{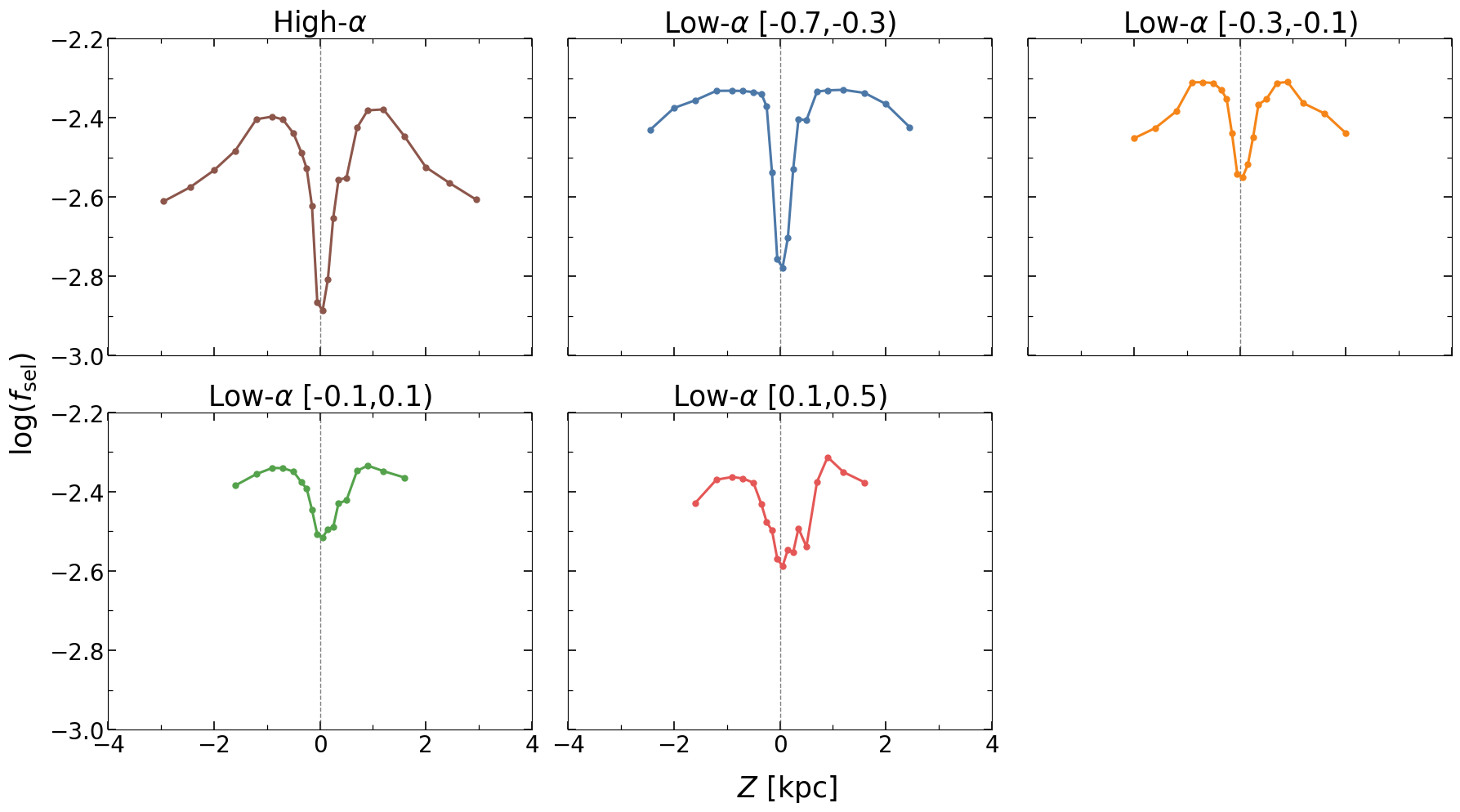}
    \caption{
    Selection functions $f_{\rm sel}$ of different MAPs in the spatial cell defined by $-9.2\leq X<-8.7\,{\rm kpc}$ and $-2.5\leq Y<-2.0\, {\rm kpc}$, \rev{corresponding to a Galactocentric radius of $R\simeq9.2\,{\rm kpc}$.}
    }
    \label{fig:selection_function}
    \end{figure*}
    
    \begin{figure*}[ht!]
    \sidecaption
    \includegraphics[width=12cm]{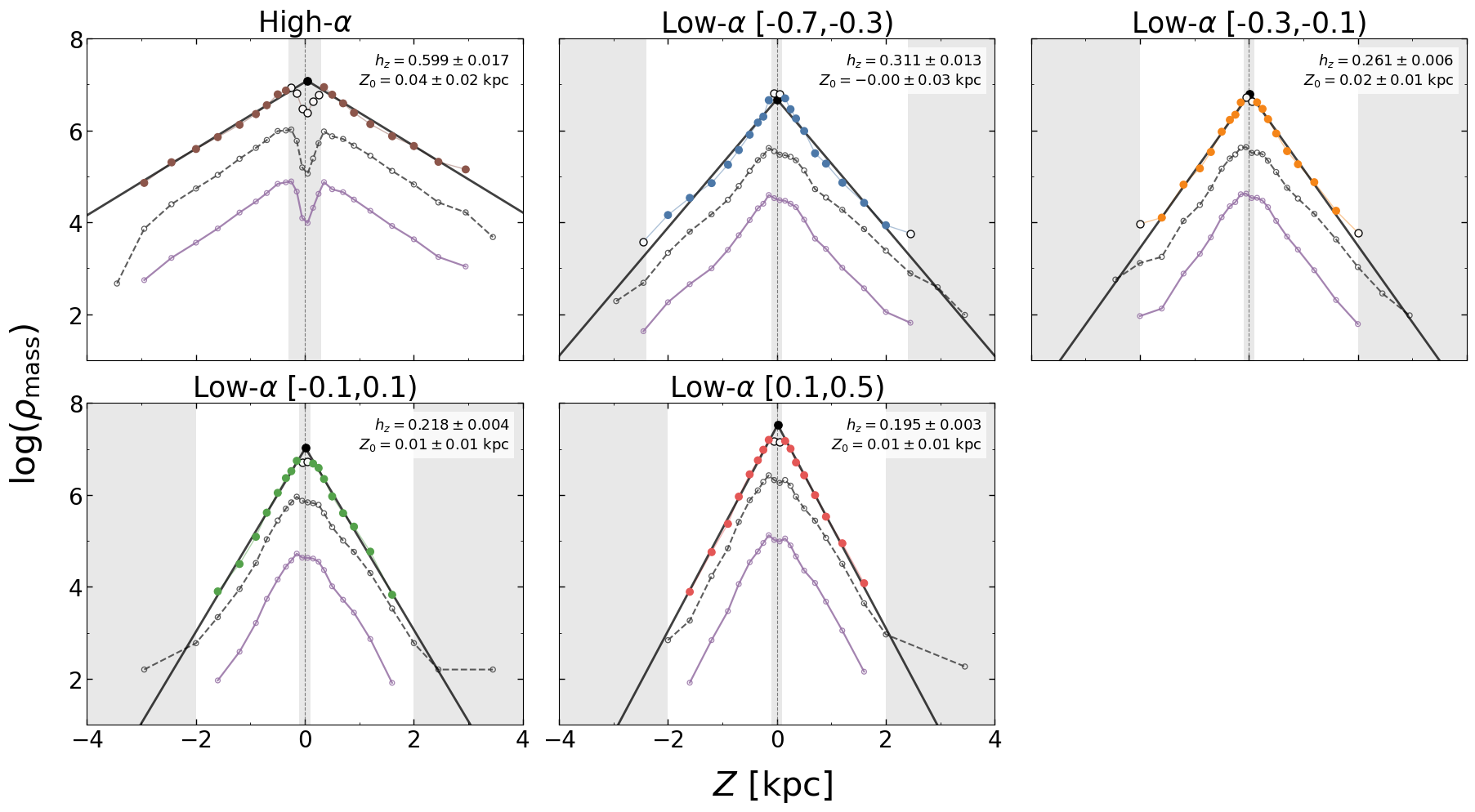}
    \caption{
    Example of the vertical density-profile fitting for different MAPs in the same spatial cell shown in Fig.~\ref{fig:selection_function}. The colored points show the observed vertical stellar mass density profiles as a function of height $Z$, and the solid black lines show the best-fitting exponential models. The gray shaded regions indicate the masked ranges excluded from the fitting procedure, while the open circles denote the corresponding masked data points. The dashed vertical lines mark the location of the Galactic mid-plane ($Z=0$). The gray dashed curves show the corresponding raw density profiles before selection-function correction. \rev{The pink curve shows the density profile after applying the XP completeness correction but before the selection-function correction.} These profiles are not normalized and are included only for comparison of their overall shapes.
    }
    \label{fig:fit_example}
    \end{figure*}

    Figure~\ref{fig:selection_function} shows an example of the selection function as a function of height $Z$ for different MAPs within a single $XY$ cell. Near the Galactic mid-plane, the selection function tends to be lower because of stronger extinction. At larger $|Z|$, although extinction becomes weaker, the increasing heliocentric distance makes the apparent magnitude limit of $G\leq16$ more restrictive, leading to an overall decrease in the selection function with height.
    
    After obtaining both the completeness correction and the selection function, we estimate the stellar mass density for each MAP in each spatial element. The mass density is defined as
    \begin{equation}
    \begin{aligned}
    \rho_*(X,Y,Z)
    &=
    \frac{N_{\rm obs}}{V}
    \frac{\bar{m}}{f_{\rm comp}\,f_{\rm sel}} ,
    \end{aligned}
    \label{eq:mass_density}
    \end{equation}
    where $N_{\rm obs}$ is the number of observed XP RGB stars in the spatial element, $f_{\rm comp}$ is the XP completeness derived in the previous subsection, $V$ is the volume of the spatial element, and $f_{\rm sel}$ is the corresponding selection function for the given MAP and spatial element. The quantity $\bar{m}$ represents the mean living stellar mass weighted by the age distribution and the IMF, defined as
    \begin{equation}
    \begin{aligned}
    \bar{m}
    &=
    \sum_{\tau} w_{\tau}
    \sum_j p_j\left(\tau,[\alpha/{\rm M}],[{\rm M/H}]\right)
    m_{{\rm live},j} ,
    \end{aligned}
    \label{eq:mean_mass}
    \end{equation}
    where $w_{\tau}$ is the age weight, $p_j$ is the IMF weight of the $j$ mass point on the isochrone, and $m_{{\rm live},j}$ is the corresponding living stellar mass at that age. This procedure converts the observed number of XP RGB stars into the intrinsic living stellar mass density of each MAP in each spatial element. \rev{The selection-function correction may introduce additional systematic uncertainties into the reconstructed stellar mass densities, mainly through the adopted extinction correction and stellar evolutionary models. The uncertainty associated with extinction is mitigated in the subsequent analysis by masking regions with high extinction. \citet{Lian2022a} tested the impact of the adopted stellar evolutionary models by replacing the PARSEC isochrones with MIST isochrones and found that the resulting structural parameters generally differ by less than 10\%, while the main structural trends remain unchanged. This suggests that the systematic uncertainty associated with the choice of stellar evolutionary models is unlikely to significantly affect the main structural trends discussed in this work.}

    \subsection{Scale-height fitting}
    After applying the completeness correction and the selection-function correction, the vertical stellar mass-density profiles of different MAPs generally show clear exponential declines. Therefore, for each MAP and each spatial cell in the $XY$ plane, we fit the vertical mass-density profile with a single-exponential model,
    \begin{equation}
    \rho(Z)=\rho_0\exp\left(-\frac{|Z-Z_0|}{h_z}\right),
    \label{eq:scale_height_model}
    \end{equation}
    where $\rho_0$ is the mass density at $Z=Z_0$, $Z_0$ is the fitted mid-plane offset, and $h_z$ is the scale height of the corresponding MAP in that spatial cell.

    \rev{The Galactic warp can introduce systematic vertical displacements of the disk and, if not properly accounted for, may affect the inferred scale heights and flaring trends \citep{Poggio2026,Khanna2025}. In our fitting, $Z_0$ is treated as a free parameter rather than being fixed at $Z=0$, allowing variations in the local mid-plane position to be partially absorbed by the fitted density profile. This treatment therefore reduces the potential impact of warp-induced vertical displacements on the inferred scale heights.}
    
    To reduce the impact of small-number statistics on the fitting results, we impose a minimum star-count requirement on the spatial elements used in the fitting. For all stellar populations considered in this work, we require that each spatial element contains at least $N_{\rm obs}\geq5$ stars satisfying the same observational selection criteria used in the selection-function calculation. Spatial elements below this threshold are excluded from the scale-height fitting. \rev{We have also tested a more conservative threshold of $N_{\rm obs}\geq20$ and find that our main results and conclusions remain unchanged.}

    \begin{figure*}[ht!]
    \sidecaption
    \includegraphics[width=12cm]{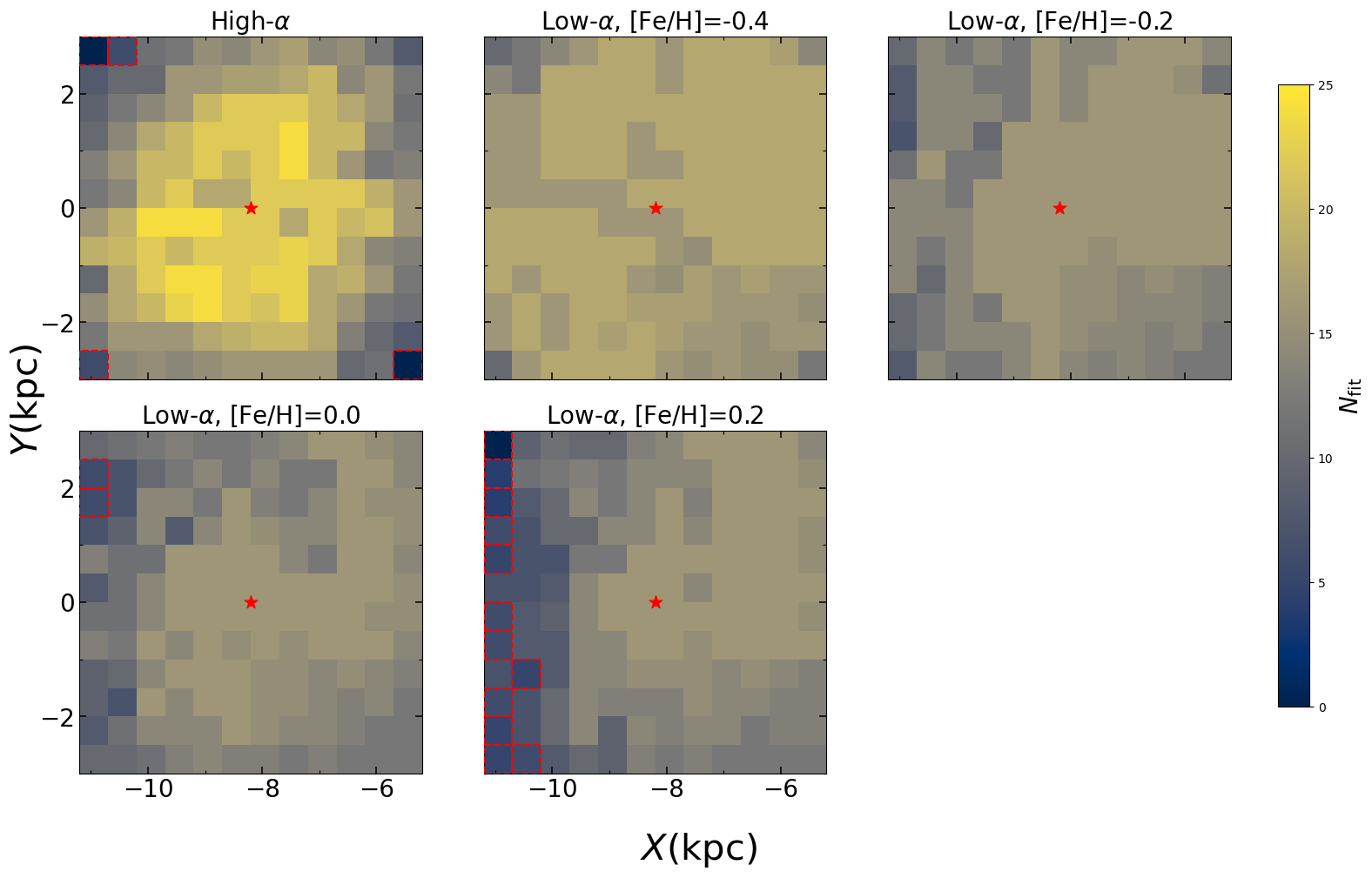}
      \caption{Distribution of the number of fitted points ($N_{\mathrm{fit}}$) in the Galactic $XY$ plane for different MAPs. The color scale indicates the number of points used in the vertical density-profile fitting within each spatial bin. Red dashed boxes denote bins excluded from the subsequent analysis because of insufficient fitted points. The red star marks the solar position.}
         \label{figN}
    \end{figure*}
    
    Near the Galactic mid-plane, although the selection function already accounts for the effects of extinction, the corrected mass densities in some highly extincted regions may still show large deviations from a simple exponential profile. These deviations are most likely caused by uncertainties in the extinction correction. To avoid these outlying points affecting the scale-height measurements, we symmetrically mask the central region for each MAP and require all MAPs to exclude at least the region of $|Z|<0.1\,{\rm kpc}$. For spatial cells in which the central profile shows clear flattening or depression, a wider symmetric mask is adopted. \rev{For the high-$\alpha$ sample, a small number of spatial elements show strong deviations from a simple exponential profile even in the uncorrected number-density distributions. To prevent these localized outliers from affecting the scale-height measurements, the corresponding points are also symmetrically masked.}
    
    In addition, different MAPs have different vertical extents. To avoid possible contaminating points far from the disk plane affecting the fitting results, different fitting ranges in height are adopted for different MAPs. For the high-$\alpha$ population, we use the full available vertical range after applying the masks and star-count criteria. The low-$\alpha$ populations are generally more concentrated toward the Galactic disk than the high-$\alpha$ sample. For the metal-poor low-$\alpha$ population, which is the most vertically extended among the low-$\alpha$ MAPs \citep{Lian2022a,Imig2023}, we only use points within $|Z|<2.4\,{\rm kpc}$ for the fitting. For the other low-$\alpha$ populations, whose densities decline more rapidly with height, we only fit points within $|Z|<2.0\,{\rm kpc}$.
    
    Figure~\ref{fig:fit_example} shows an example of the vertical stellar mass-density profiles and the corresponding best-fitting results for different MAPs in a representative $XY$ spatial cell. The solid black lines show the best-fitting exponential models, the gray shaded regions indicate the height ranges masked in the fitting, and the open circles denote the excluded spatial elements.

    \begin{figure*}[t]
    \centering
    \includegraphics[width=1.0\textwidth]{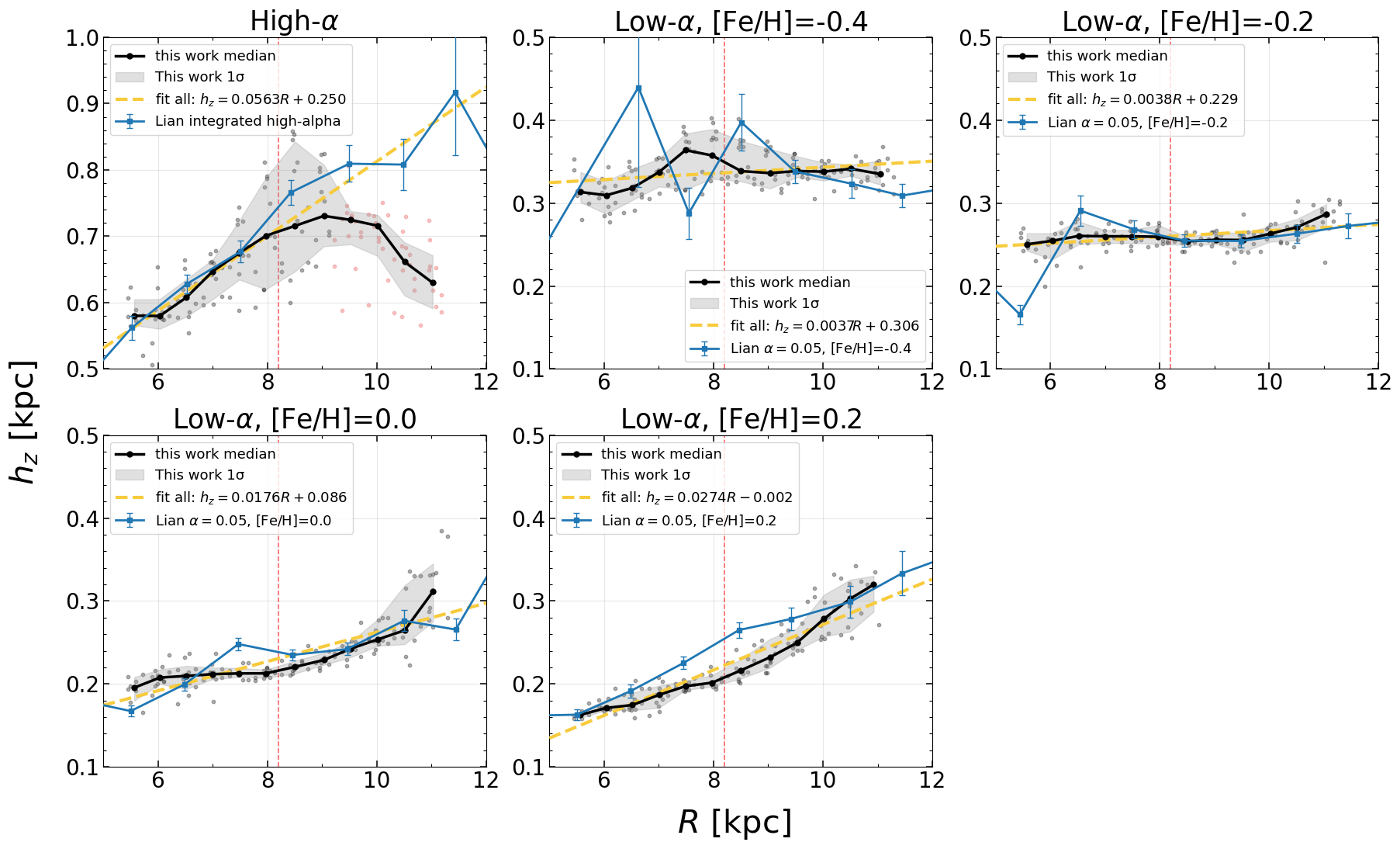}
      \caption{Radial variation of the scale height $h_z$ for different MAPs. In each panel, the gray points show the scale heights measured in individual spatial bins at different Galactocentric radii. The black solid line represents the median relation, and the gray shaded region indicates the 16th–84th percentile range. Gray points indicate bins included in the fitting, while red points denote bins excluded by the masking procedure. The yellow lines show the best-fitting linear relations to the individual measurements, with the corresponding fitting results listed in the legends. The blue line and error bars show the corresponding measurements from \citet{Lian2025} for the closest matching MAP.}
         \label{fighr}
    \end{figure*}
   
    To ensure the reliability of the fitted scale heights in each spatial cell, we further impose a minimum requirement on the number of data points used in the fitting. As shown in Fig.~\ref{figN}, for each MAP and each $XY$ cell, only cells with more than six unmasked data points after applying all masks are retained as valid cells in the final analysis.

\section{Results and Discussion}

    \subsection{Radial variation of scale height}
    Figure~\ref{fighr} shows the radial variation of the scale height $h_z$ for different MAPs. The black solid lines indicate the median scale-height profiles derived in this work, while the gray shaded regions show the corresponding scatter. The yellow solid lines show the linear fits to the scale-height profiles measured in this work, which are used to quantify the flaring strength of different MAPs.
    
    For the high-$\alpha$ population, the outer disk region may be affected by contamination. As shown by the $[\alpha/{\rm M}]$–$[{\rm M/H}]$ distributions of different spatial cells in Fig.~\ref{fig:chemical_xy}, the Gaia XP high-$\alpha$ selection increasingly deviates from the high-resolution APOGEE abundance distribution at large Galactocentric radii. In the APOGEE sample, the genuine high-$\alpha$ sequence declines rapidly with radius and becomes very weak in the outer disk. In contrast, a substantial number of stars are still classified as high-$\alpha$ in the Gaia XP sample at large radii. Since metal-poor low-$\alpha$ stars dominate the high-$|Z|$ stellar population in the outer disk, abundance uncertainties in Gaia XP may lead to contamination of the high-$\alpha$ sample. As a result, the outer-disk high-$\alpha$ sample may not fully represent the genuine high-$\alpha$ sequence. Therefore, when fitting the radial variation of the high-$\alpha$ scale height, we exclude the region at $R>9.2\,{\rm kpc}$ to minimize the impact of outer-disk contamination on the inferred flaring strength.
    
    The linear fitting results show that all MAPs exhibit some degree of flaring, with their scale heights generally increasing with Galactocentric radius. The high-$\alpha$ population shows the strongest flaring, followed by the metal-rich and solar-metallicity low-$\alpha$ populations. The metal-poor and sub-solar low-$\alpha$ populations exhibit the weakest flaring, with only a small difference between their fitted radial gradients. In particular, the solar-metallicity low-$\alpha$ population shows only weak flaring over the range $5<R<8\,{\rm kpc}$, followed by a more prominent increase in scale height at $R\gtrsim8\,{\rm kpc}$.

    In Figure~\ref{fighr}, the blue solid lines and error bars represent the results of the corresponding MAPs from \citet{Lian2025}, which are based on APOGEE data. Overall, the scale-height profiles obtained in this work agree well with those of \citet{Lian2025} in both shape and normalization. Flaring of chemically defined MAPs has also been discussed in earlier APOGEE studies. Using APOGEE red-clump stars, \citet{Bovy2016} found that low-$\alpha$ MAPs flare, although their high-$\alpha$ MAPs show nearly constant scale heights. The discrepancy in the high-$\alpha$ population may be partly related to differences in sample selection and in the treatment of the selection function. Similar flaring trends have also been reported for mono-age populations of LAMOST red-clump stars by \citet{Yu2025}, consistent with the theoretical expectation that mono-age populations generally flare in inside-out disk formation scenarios \citep{Minchev2015}. From an external-galaxy perspective, \citet{Chen2026} found that local disk galaxies are generally thin and show negligible flaring when their integrated light distributions are fitted with a single-component disk model. They noted that the measured scale heights and radial gradients are consistent with those of the Milky Way geometric thin disk, which is represented by metal-rich low-$\alpha$ populations. This appears to be different from our result. The difference may partly arise from the different radial ranges probed by the two studies. \citet{Chen2026} measured the scale-height variation within $1.5\,R_{\rm e}$ of external galaxies, whereas our analysis probes the Galactic disk over a broader radial range of approximately $1$--$2\,R_{\rm e}$. Together, these studies show that flaring is a common feature of disk sub-populations, while our results and those of \citet{Lian2025} further indicate that the flaring strength varies among different chemical populations.

    For a stellar population with a given vertical velocity dispersion, the decrease in the disk surface density with Galactocentric radius weakens the vertical restoring force in the outer disk. As a result, stars can reach larger vertical heights, leading to a larger measured scale height. In this work, all MAPs show different degrees of flaring, indicating that the weakening of the vertical gravitational potential in the outer disk is a common effect among different chemical populations. However, the flaring strengths differ significantly from one MAP to another, suggesting that the radial variation of the gravitational potential alone is not sufficient to fully explain the differences in their vertical structures. Other factors, including the thickness of the gas disk at the time of star formation, birth velocity dispersion, radial migration, external perturbations, and long-term dynamical heating, may also contribute to the observed flaring strengths.

    For the high-$\alpha$ population, the stars are generally old, dynamically hot, and vertically extended, and are therefore commonly associated with the early thick-disk component of the Milky Way \citep{Fuhrmann1998,Reddy2006,Haywood2013,Pinna2019}. Recent observations of high-redshift edge-on disk galaxies also support this picture. Based on \rev{data from the James Webb Space Telescope COSMOS-Web survey}, \citet{Yu2026} found that high-redshift disk galaxies are generally characterized by intermediate vertical thickness rather than by a well-developed thin--thick disk structure. Similar results have also been reported by \citet{Hamilton2023} and \citet{Lian2024b}, who found that disk galaxies at high redshift are substantially thicker than their local counterparts. Together, these observations suggest that the progenitors of present-day thick disks were already vertically extended and dynamically hot. The fact that the high-$\alpha$ population exhibits the strongest flare is consistent with its hot dynamical nature and early formation history. One possible explanation is that the gas disk in the early Milky Way was more turbulent and vertically thicker, so that high-$\alpha$ stars were born with intrinsically larger vertical heights \citep{Brook2004,Bournaud2009}. Another possibility is that the high-$\alpha$ population has experienced a longer period of dynamical heating and radial migration. In this case, some high-$\alpha$ stars that formed in the inner disk may have migrated or been dynamically perturbed to larger radii, where the weaker vertical potential allows them to reach larger vertical amplitudes and hence show larger scale heights \citep{Sellwood2002,Schoenrich2009,Minchev2011,Solway2012,VeraCiro2014}.

    For the low-$\alpha$ populations, the differences in flaring strength among different MAPs are unlikely to be caused by a single mechanism. Instead, they may reflect the combined effects of dynamical heating, radial redistribution, and external perturbations. Older low-$\alpha$ populations are generally expected to have larger vertical velocity dispersions due to age-dependent heating \citep{Wielen1977,Nordstrom2004}. If such dynamically hotter stars are redistributed to larger radii, the weaker vertical restoring force in the outer disk can translate the larger vertical action into larger vertical amplitudes, producing a stronger flaring signature. In this sense, radial migration does not by itself guarantee stronger flaring, but it can amplify the flaring of populations that are already vertically hotter.
    
    Second, external perturbations may also modify the vertical structure of the gas disk and newly formed stars at specific epochs. Gaia DR2 revealed a clear phase spiral in the Galactic disk, indicating that the Milky Way disk is still dynamically perturbed and undergoing phase mixing \citep{Purcell2011,Gomez2013,Antoja2018,Laporte2019,Bennett&Bovy2019,Li2021}. The Sagittarius dwarf galaxy has been suggested as one of the possible perturbers responsible for such non-equilibrium features \citep{Laporte2019}. In addition, \citet{RuizLara2020} reconstructed the star formation history within about $2\,{\rm kpc}$ of the Sun using Gaia DR2 and found three prominent episodes of enhanced star formation at approximately 5.7, 1.9, and $1.0\,{\rm Gyr}$ ago. These episodes are temporally close to proposed pericentric passages of Sagittarius, suggesting that Sagittarius may have repeatedly affected the star formation history of the Galactic disk.
    
    In this context, the flaring behavior of the solar-metallicity low-$\alpha$ population may be particularly informative. This population shows a clear increase in scale height at $R\gtrsim8\,{\rm kpc}$. One possible interpretation is that, because the outer disk has a weaker vertical restoring force, perturbations from satellite galaxies such as Sagittarius can more easily produce significant vertical thickening there. If the main formation epoch of the solar-metallicity low-$\alpha$ population was close to a phase of Sagittarius-induced perturbation, Sagittarius may have disturbed the gas disk, enhanced gas turbulence, or triggered star formation in the outer disk. Stars formed during this period may therefore have been born in a thicker or dynamically hotter gas disk, producing a stronger flare in the outer disk. In contrast, the metal-poor low-$\alpha$ and sub-solar low-$\alpha$ populations show weaker flaring, suggesting that their vertical structures vary more smoothly over the radial range studied here and may not have been affected by the same combination of processes to the same extent.
    
    Nevertheless, the current scale-height measurements alone cannot uniquely determine the contribution of Sagittarius to the flaring of different MAPs, nor can they quantitatively separate the effects of birth thickness, long-term dynamical heating, radial migration, and external perturbations. More precise age estimates, kinematic information, and orbital properties will be required to further disentangle the contributions from different mechanisms.

    \subsection{Radial variation of surface mass density}
\begin{figure*}[ht!]
    \centering
    \includegraphics[width=\hsize]{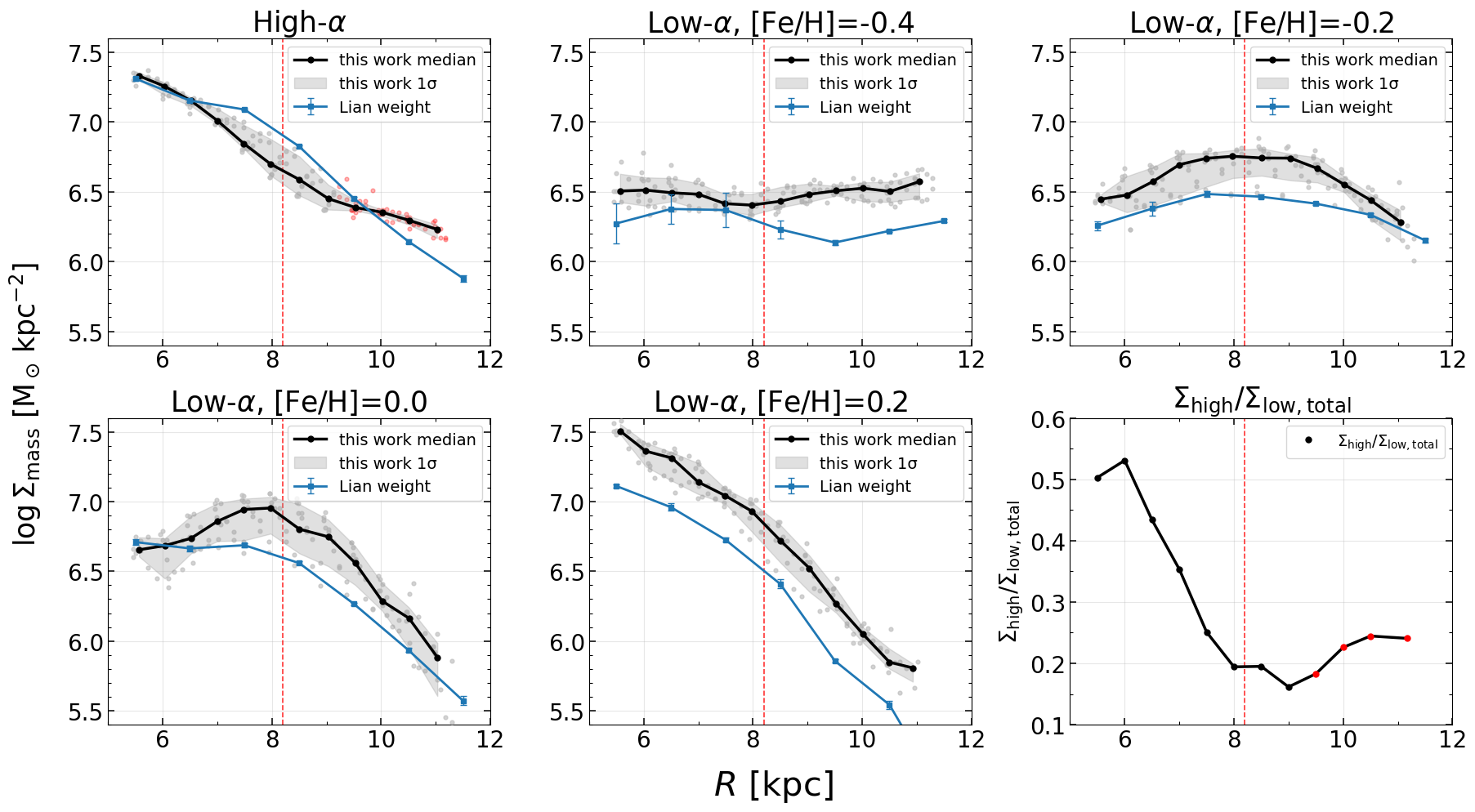}
      \caption{Radial variation of the surface-mass-density $\Sigma_{\mathrm{mass}}$ for different MAPs. Gray points connected by lines show the surface mass densities derived for individual spatial bins by integrating the best-fitting vertical density profiles. The black solid line represents the median relation, while the gray shaded region indicates the 16th–84th percentile range. The blue solid lines and error bars show the corresponding APOGEE-based surface-mass-density profiles from \citet{Lian2024}. The vertical red dashed line marks the solar radius ($R=8.2\,\mathrm{kpc}$). \rev{The bottom-right panel shows the radial variation of the surface-mass-density ratio between the high-$\alpha$ population and the combined low-$\alpha$ populations. As in Fig.~\ref{fighr}, the red points indicate the potentially contaminated outer-disk region at $R>9.2\,\mathrm{kpc}$ for the high-$\alpha$ population.}}
         \label{density}
    \end{figure*}

    In addition to the vertical scale height, the radial surface-mass-density profile provides an important constraint on the radial mass distribution and assembly history of different disk populations. Figure~\ref{density} shows the radial variation of the surface-mass-density $\Sigma_{\rm mass}$ for different MAPs. The surface mass densities are obtained by integrating the best-fitting vertical density profiles described above. The black solid lines indicate the median surface-mass-density profiles derived in this work, while the gray shaded regions show the corresponding scatter.

    The surface-mass-density profiles reveal clear differences \jl{among} different MAPs. The high-$\alpha$ and metal-rich low-$\alpha$ populations show approximately exponential declines. The metal-poor low-$\alpha$ population exhibits a much flatter profile. The sub-solar low-$\alpha$ population shows a mildly positive slope in the inner disk, followed by a relatively flat plateau and then a transition to a negative radial gradient at larger radii. The solar-metallicity low-$\alpha$ population is nearly flat in the inner disk, showing a short rising segment, followed by a clearer exponential decline beyond $R\simeq7.5\,{\rm kpc}$. \rev{To quantify the approximately exponential profiles of the high-$\alpha$ and metal-rich low-$\alpha$ populations, we perform simple single-exponential fits of the form $\Sigma_{\rm mass}(R)\propto\exp(-R/h_R)$ over the radial range covered by our analysis. For the high-$\alpha$ population, we exclude the potentially contaminated outer-disk region at $R>9.2\,\mathrm{kpc}$ from the fit, as discussed above. We obtain radial scale lengths of $h_R=1.632\,\mathrm{kpc}$ for the high-$\alpha$ population and $h_R=1.297\,\mathrm{kpc}$ for the metal-rich low-$\alpha$ population. These scale lengths are broadly consistent with previous measurements of the radial structure of the high-$\alpha$ and metal-rich low-$\alpha$ populations \citep{Lian2022a,Khoperskov2025}.}
    
    In Figure~\ref{density}, the blue solid lines and error bars represent the APOGEE-based results from \citet{Lian2024}. Since the low-$\alpha$ MAPs in \citet{Lian2024} are defined with finer chemical-abundance bins than those adopted in this work, we calculate the fractional overlap in abundance space between their MAP bins and our low-$\alpha$ bins. These overlap fractions are then used as weights to obtain the weighted APOGEE profiles shown in the figure. Overall, the surface-mass-density profiles derived in this work agree well with those of \citet{Lian2024} in terms of their radial shapes. For all populations, the surface mass densities derived in this work are generally higher than those from \citet{Lian2024}. The origin of this systematic offset is unclear, but it may partly reflect differences in the survey strategies of Gaia XP and APOGEE, as well as other systematic differences between the two analyses. We therefore focus mainly on the radial shapes of the profiles rather than on their absolute normalizations. For the high-$\alpha$ population, as discussed in the previous section and also suggested by the chemical-abundance distributions of different spatial cells shown in Fig.~\ref{fig:chemical_xy}, the Gaia XP high-$\alpha$ sample may suffer from significant contamination in the outer disk at $R\gtrsim9\,{\rm kpc}$. As a result, the high-$\alpha$ surface-mass-density profile shows some deviation from the APOGEE result at large Galactocentric radii. \rev{The bottom-right panel of Fig.~\ref{density} shows the radial variation of the surface-mass-density ratio between the high-$\alpha$ population and the combined low-$\alpha$ populations. The relative contribution of the high-$\alpha$ population decreases with increasing Galactocentric radius. At larger radii, the trend becomes flatter, which is likely related to the increasing contamination of the high-$\alpha$ sample discussed above. We caution that our adopted chemical abundance bins do not provide a continuous partition of the high- and low-$\alpha$ sequences, with a gap between the two populations. Therefore, the derived mass ratio should be regarded as indicative of the relative contributions of the populations defined in this work, rather than as a complete decomposition of the Galactic disk into high- and low-$\alpha$ components, and should not be directly compared quantitatively with mass ratios derived using different chemical definitions.}

    \jl{The high-$\alpha$ and metal-rich low-$\alpha$ populations both show relatively steep, nearly exponential declines, suggesting a more concentrated inner-disk origin. The high-$\alpha$ population is generally associated with the early, rapidly formed thick-disk component, while the metal-rich low-$\alpha$ population may have formed subsequently from gas that had already been highly enriched in the inner disk. Therefore, although these two populations occupy different sequences in chemical abundance space, they may still retain a similar inner-disk origin in their spatial distributions. This result is consistent with the evolutionary connection between the high-$\alpha$ and metal-rich low-$\alpha$ populations proposed by \citet{Lian2020c}, and is also broadly consistent with the discussion in \citet{Lian2022a} that these populations are closely related to the inner-disk structure.} 

    The non-exponential radial profiles of the low-$\alpha$ MAPs provide an important clue to the radial assembly of the Galactic disk. In contrast to the high-$\alpha$ and metal-rich low-$\alpha$ populations, which approximately follow declining radial profiles, the other low-$\alpha$ MAPs show plateaus, positive radial gradients, or clear radial breaks in their surface-mass-density profiles. Such structures are qualitatively consistent with the broken luminosity surface-density profile reported by \citet{Lian2024}, who found a nearly flat disk component between $R\simeq3.5$ and $7.5\,{\rm kpc}$, followed by an exponential decline at larger radii. The presence of these plateau-like and broken profiles indicates that the Galactic disk cannot be fully described by a single exponential component. Instead, the radial mass distribution appears to depend strongly on stellar population, suggesting that different MAPs preserve different stages of disk growth. In particular, the more extended distributions and inner rising profiles of some younger low-$\alpha$ populations are naturally consistent with the size growth and inside-out formation of the Galactic disk. A similar plateau-like radial structure has also been reported by \citet{Wu2026} using a large sample of Gaia XP red-clump stars. They found that the radial density profile of the Galactic disk is not well described by a single exponential profile, but instead consists of several radial components. Specifically, their derived profile shows a nearly flat plateau at $R\sim3{-}7\,{\rm kpc}$, an exponential decline from beyond the solar radius to $R\sim13\,{\rm kpc}$, and a sharper exponential drop-off at $R\gtrsim13\,{\rm kpc}$. This is qualitatively consistent with the plateau-like features found in some of our low-$\alpha$ surface-mass-density profiles, and further supports the view that the radial mass distribution of the Galactic disk is more complex than a simple exponential disk.
    
    In particular, the solar-metallicity low-$\alpha$ population is relatively flat in the inner disk and changes to a clearer exponential decline beyond $R\simeq8\,{\rm kpc}$. This transition radius is close to the radius at which the total surface-density profile in \citet{Lian2024} changes from an inner plateau to an outer exponential decline, suggesting that the solar radius may lie near an important transition region in the radial structure of the Milky Way disk. This region may be affected simultaneously by inside-out growth, radial migration, bar- or spiral arm-driven dynamical redistribution, and variations in the star formation efficiency of the outer disk.

    \rev{Recent studies suggest that plateau or inner broken profiles in disk galaxies may arise through multiple physical pathways, with stellar-population evolution and dynamical redistribution representing two possible mechanisms \citep{Zhao2026}. One possible pathway is central homogeneous quenching. In barred galaxies, bars can redistribute gas and alter the spatial distribution and efficiency of star formation through gas flows, angular-momentum redistribution, shear, and turbulence \citep{Wozniak2007,James2009,Fraser2020}. If star formation is suppressed over an extended inner region while remaining active in the surrounding disk, the inner stellar population gradually becomes older and redder, producing strong radial gradients in stellar age and color. Such stellar-population gradients can generate a plateau in the luminosity profile, with a stronger effect at bluer wavelengths because blue-band emission is more sensitive to recent star formation \citep{Nersesian2020,Zhao2026}. Therefore, plateaus produced predominantly through this pathway are expected to exhibit significant wavelength dependence.}

    \rev{Another possible pathway is dynamical redistribution. Non-axisymmetric structures such as bars and spiral arms can exchange angular momentum with stars and drive radial migration, thereby directly reshaping the stellar mass distribution \citep{Sellwood2002,Minchev2010,Thor2021,RuizLara2017}. Numerical
    simulations have shown that bar-driven stellar redistribution can transform an initially exponential stellar mass surface-density profile into a broken profile with a flattened inner region \citep{Thor2021}. Because this mechanism directly modifies the underlying stellar mass distribution, a plateau produced predominantly by dynamical redistribution is expected to exhibit relatively weak wavelength dependence and weaker stellar-population gradients \citep{Zhao2026}. These two mechanisms are not necessarily mutually exclusive, and their relative importance may vary among galaxies. In the Milky Way, the dependence of the plateau on stellar population age \citep{Lian2024}, together with its persistence in the stellar mass surface-density profile \citep{Lian2025b}, suggests that both stellar-population evolution and dynamical redistribution may contribute to the observed radial structure.}

\section{Summary}
    This work investigates the structural properties of different MAPs in the Galactic disk using the Gaia XP sample of \citet{Li2024}. By correcting the XP sample completeness with a Gaia photometric parent sample and constructing selection functions based on PARSEC isochrones and the Kroupa IMF, we convert the observed stellar counts into intrinsic stellar mass densities and reconstruct the spatial distributions of different chemical populations.
    
    All MAPs exhibit varying degrees of flaring. Significant differences in flaring strength are found among the broader chemical populations, with the high-$\alpha$ population showing the strongest flaring. Among the low-$\alpha$ MAPs, the metal-rich and solar-metallicity populations show stronger radial increases in scale height, whereas the metal-poor and sub-solar populations exhibit similarly weak flaring. For the high-$\alpha$ population, the strong flaring may be related to the turbulent gaseous environment of the early Milky Way, long-term dynamical heating, and radial migration. Within the low-$\alpha$ sequence, the differences in flaring strength are likely produced by a combination of multiple mechanisms. In particular, the solar-metallicity low-$\alpha$ population shows stronger flaring than the metal-poor low-$\alpha$ and sub-solar low-$\alpha$ populations at $R\gtrsim8\,{\rm kpc}$. This behavior may be associated with pericentric passages of the Sagittarius dwarf galaxy, which could have perturbed the gas disk, enhanced gas turbulence, or triggered star formation in the outer disk, causing the outer gaseous disk to become modestly thicker at large radii. Stars formed in these regions would then inherit larger birth scale heights, resulting in enhanced flaring.

    The radial surface-mass-density profiles differ substantially among different MAPs. The high-$\alpha$ and metal-rich low-$\alpha$ populations exhibit approximately exponential declines, whereas the other low-$\alpha$ MAPs show non-exponential features such as plateaus and positive radial gradients, indicating that the Galactic disk cannot be described by a simple single-exponential disk. The similar radial structures of the high-$\alpha$ and metal-rich low-$\alpha$ populations support a common inner-disk origin, while the more radially extended distributions of the younger low-$\alpha$ populations are consistent with the inside-out growth scenario of the Galactic disk. \rev{The observed plateau structures may reflect the combined effects of stellar-population evolution associated with the radial star-formation history and dynamical redistribution driven by non-axisymmetric structures such as the Galactic bar and spiral arms.}

    The scale-height and surface-mass-density profiles derived from Gaia XP are generally consistent with those obtained from APOGEE for the corresponding MAPs. This agreement demonstrates that, after appropriate completeness and selection-function corrections, Gaia XP can reliably recover the spatial structures of chemically distinct stellar populations in the Galactic disk, providing \jl{deep insights into the Galactic structure evolution based on a more massive sample than traditional spectroscopic surveys.}
    
    Future studies combining more precise stellar ages, orbital parameters, and high-precision chemical abundances will help disentangle the relative contributions of star-formation history, radial migration, dynamical heating, and external perturbations in shaping the Galactic disk. In particular, upcoming and ongoing massive spectroscopic surveys, such as 4-meter Multi-Object Spectroscopic Telescope and SDSS-V, will provide much larger samples of stars with \jl{more precise} chemical abundances, radial velocities, and stellar parameters across wide areas of the sky, offering stronger constraints on the formation and evolutionary history of the Milky Way.

\section*{Data availability}
    The stellar catalog and the radial median profiles of scale height and stellar surface mass density accompanying this work are publicly available through the National Astronomical Data Center (NADC) at \url{https://nadc.china vo.org/res/r101893/} and will also be made available at the CDS.

\begin{acknowledgements}
    We thank the referee for the constructive comments and suggestions that helped improve this manuscript. JL acknowledges support from the National Natural Science Foundation of China (No. 12473021), the National Key R\&D Program of China (No. 2024YFA1611600), Yunnan Province Science and Technology Department Grants (Nos. 202105AE160021 and 202005AB160002), the Key Laboratory of Survey Science of Yunnan Province (No. 202449CE340002), and the Start-up Fund of Yunnan University (No. CY22623101). This work has made use of data from the European Space Agency (ESA) mission Gaia (\url{https://www.cosmos.esa.int/gaia}), processed by the Gaia Data Processing and Analysis Consortium (DPAC, \url{https://www.cosmos.esa.int/web/gaia/dpac/consortium}). Funding for the DPAC has been provided by national institutions, in particular the institutions participating in the Gaia Multilateral Agreement.
\end{acknowledgements}

\bibliographystyle{aa}
\bibliography{reference}

\begin{appendix}
\onecolumn
\nolinenumbers

\section{Selection functions in all spatial cells}
\label{app:selection}
    This section presents the selection functions of different MAPs in all spatial cells used in this work. Figure~\ref{fig:appendix_sel} shows the variation of $f_{\rm sel}$ with vertical height $Z$ for each MAP. The panels are arranged according to the spatial positions of the corresponding $XY$ cells. In each panel, the black points represent the selection-function values of individual spatial elements, while the gray lines connect spatial elements within the same spatial cell to illustrate the vertical variation of the selection function.
    
    \begin{figure}[!htbp]
    \centering
    \includegraphics[width=\textwidth,height=0.80\textheight,keepaspectratio]{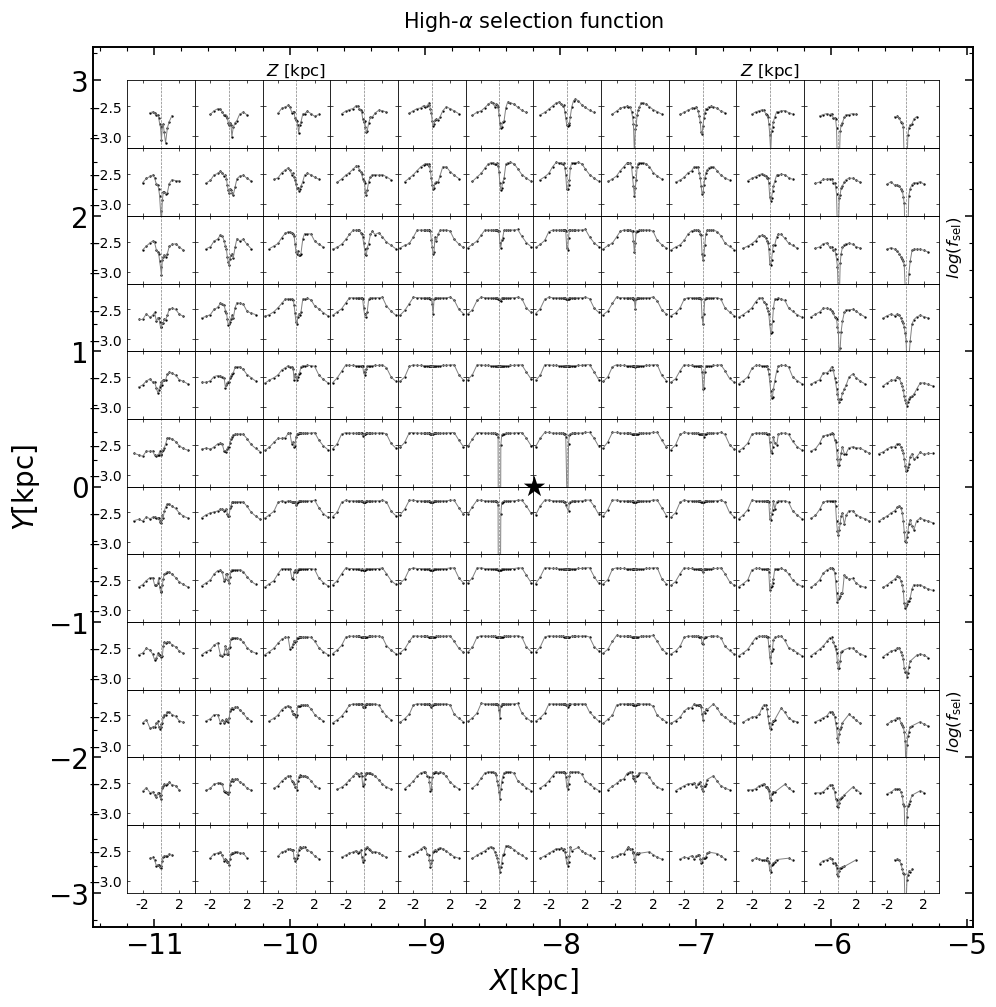}
    \caption{Selection functions $f_{\rm sel}$ as a function of height $Z$ for all spatial cells. The five continued panels show the results for different MAPs. The black points show the selection-function values in individual spatial elements. The gray solid lines connect spatial elements belonging to the same spatial cell and illustrate the corresponding selection-function profiles. The black star marks the solar position.}
    \label{fig:appendix_sel}
    \end{figure}
    
    \begin{figure}[!htbp]
    \ContinuedFloat
    \centering
    \includegraphics[width=\textwidth,height=0.80\textheight,keepaspectratio]{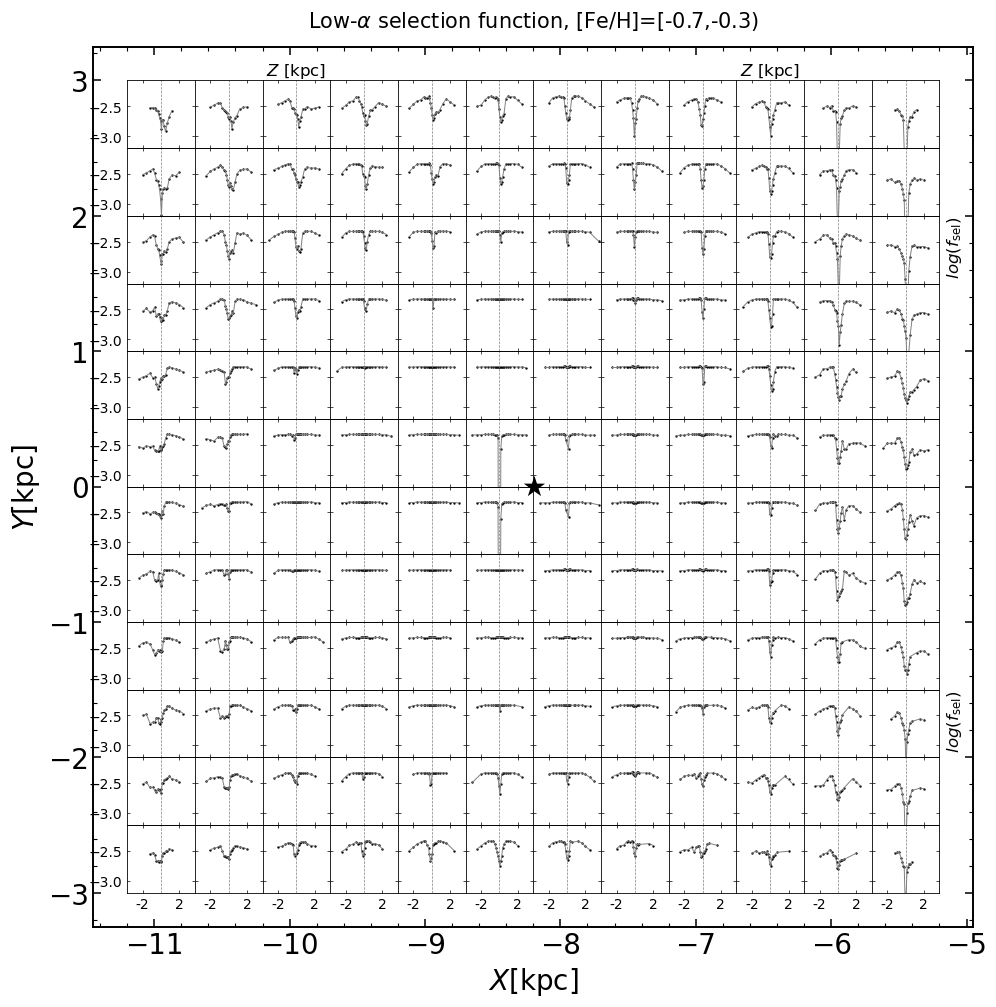}
    \caption{continued.}
    \end{figure}
    
    \begin{figure}[!htbp]
    \ContinuedFloat
    \centering
    \includegraphics[width=\textwidth,height=0.80\textheight,keepaspectratio]{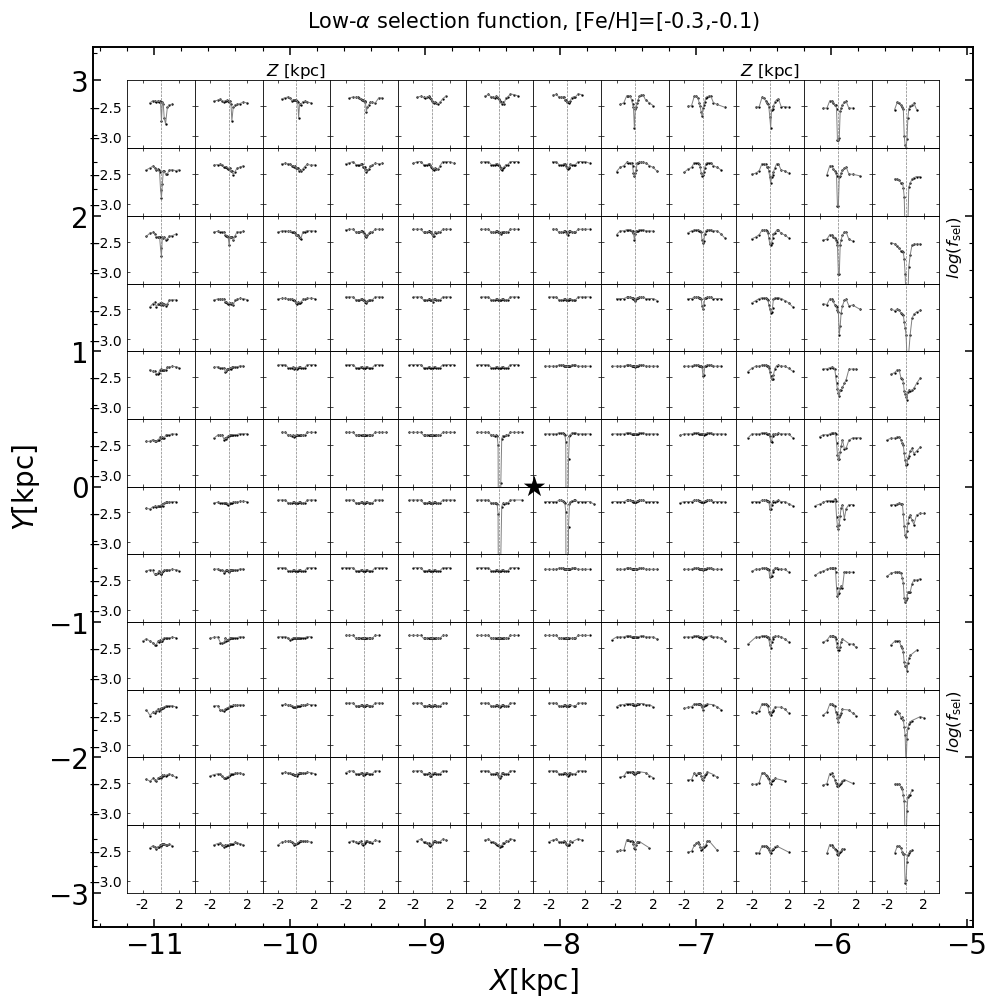}
    \caption{continued.}
    \end{figure}
    
    \begin{figure}[!htbp]
    \ContinuedFloat
    \centering
    \includegraphics[width=\textwidth,height=0.80\textheight,keepaspectratio]{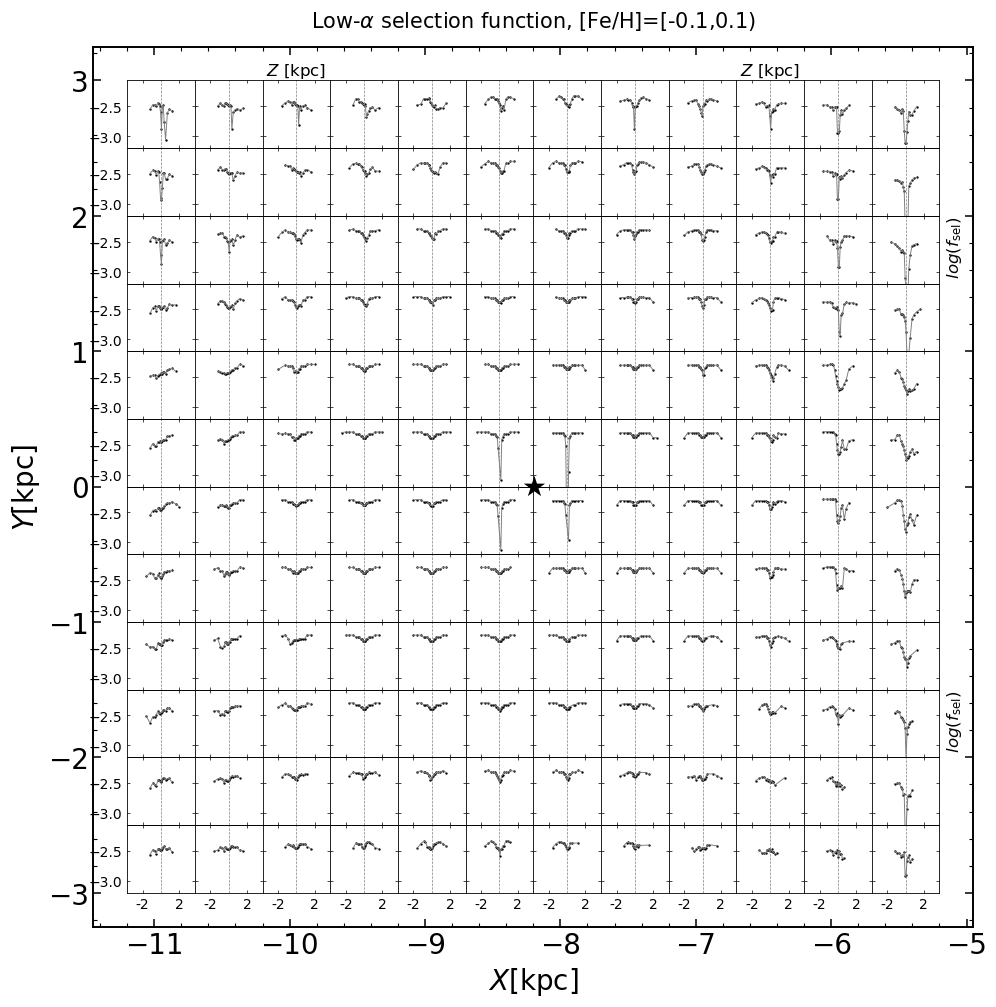}
    \caption{continued.}
    \end{figure}
    
    \begin{figure}[!htbp]
    \ContinuedFloat
    \centering
    \includegraphics[width=\textwidth,height=0.80\textheight,keepaspectratio]{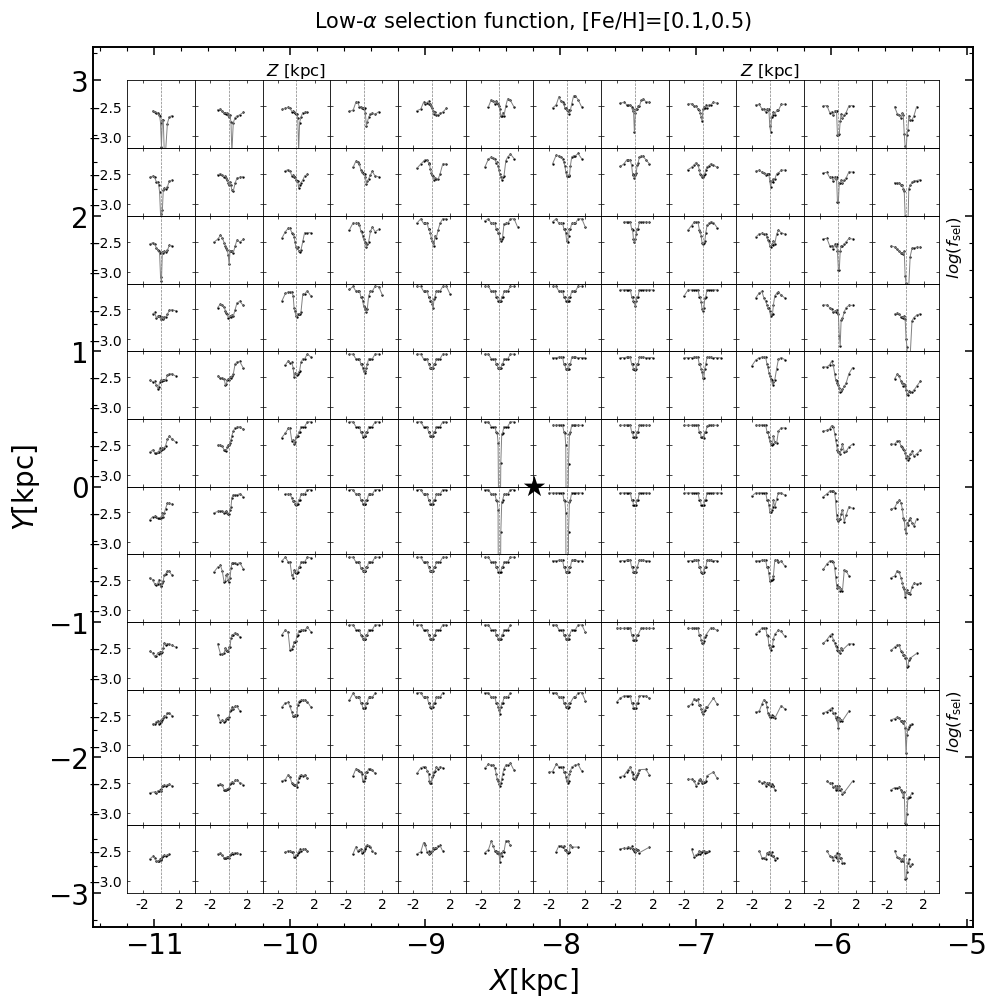}
    \caption{continued.}
    \end{figure}
    \clearpage

\section{Vertical density-profile fits in all spatial cells}
\label{app:fits}
    This section shows the vertical density-profile fitting results for all spatial cells retained in the analysis. Figure~\ref{fig:appendix_fit} presents the corrected stellar mass-density profiles of different MAPs as a function of height $Z$, together with the corresponding best-fitting exponential models. In each panel, the upper-left label gives the $X$ and $Y$ coordinate ranges of the corresponding spatial cell, while the upper-right label lists the best-fitting parameters.
    
    \begin{figure}[!htbp]
    \centering
    \includegraphics[width=\textwidth,height=0.80\textheight,keepaspectratio]{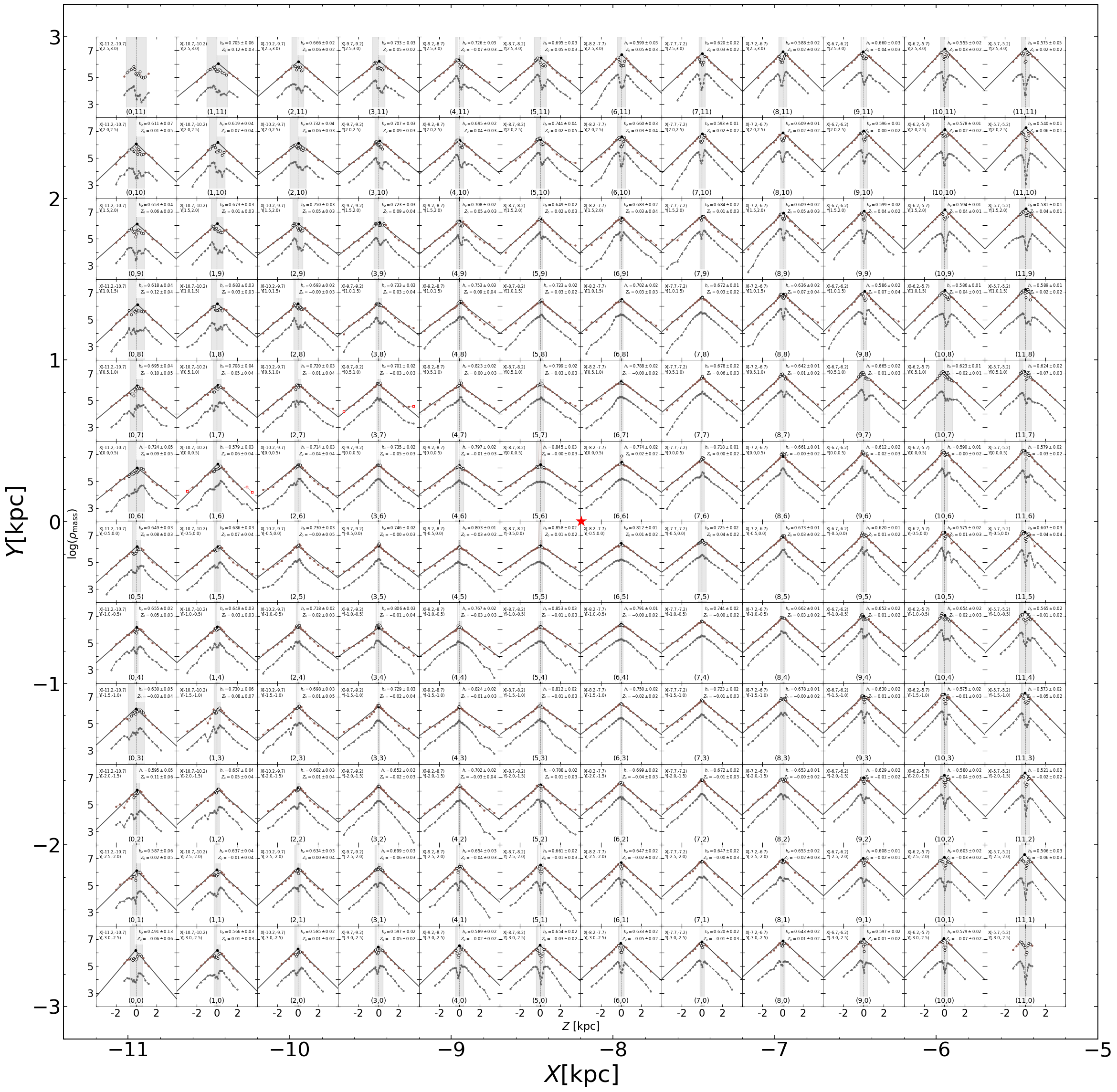}
    \caption{Vertical density-profile fitting results for all spatial cells. The five continued panels show the results for different MAPs. The colored points show the observed stellar mass density profiles as a function of height $Z$, and the solid black lines show the best-fitting exponential models. The gray shaded regions indicate the height ranges excluded from the fitting procedure, while the open circles denote the corresponding masked data points. Red open squares indicate data points that were manually and symmetrically masked. The vertical dashed line marks the location of the Galactic mid-plane ($Z=0$). The red star marks the solar position. The gray solid curves show the corresponding raw density profiles before selection-function correction. These profiles are not normalized and are included only to illustrate the overall shapes of the \jl{raw} density distributions for different MAPs.}
    \label{fig:appendix_fit}
    \end{figure}
    
    \begin{figure}[!htbp]
    \ContinuedFloat
    \centering
    \includegraphics[width=\textwidth,height=0.80\textheight,keepaspectratio]{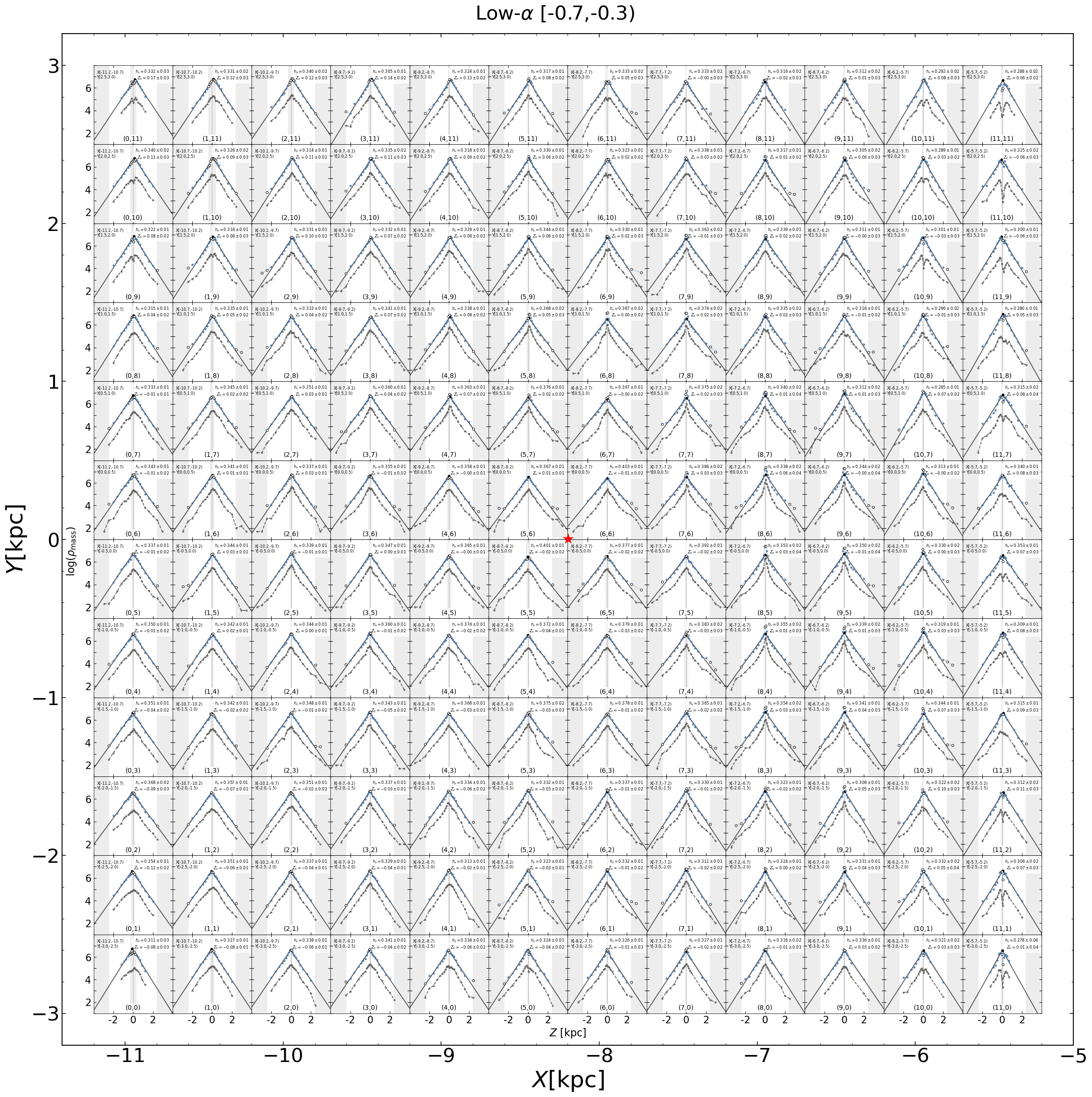}
    \caption{continued.}
    \end{figure}
    
    \begin{figure}[!htbp]
    \ContinuedFloat
    \centering
    \includegraphics[width=\textwidth,height=0.80\textheight,keepaspectratio]{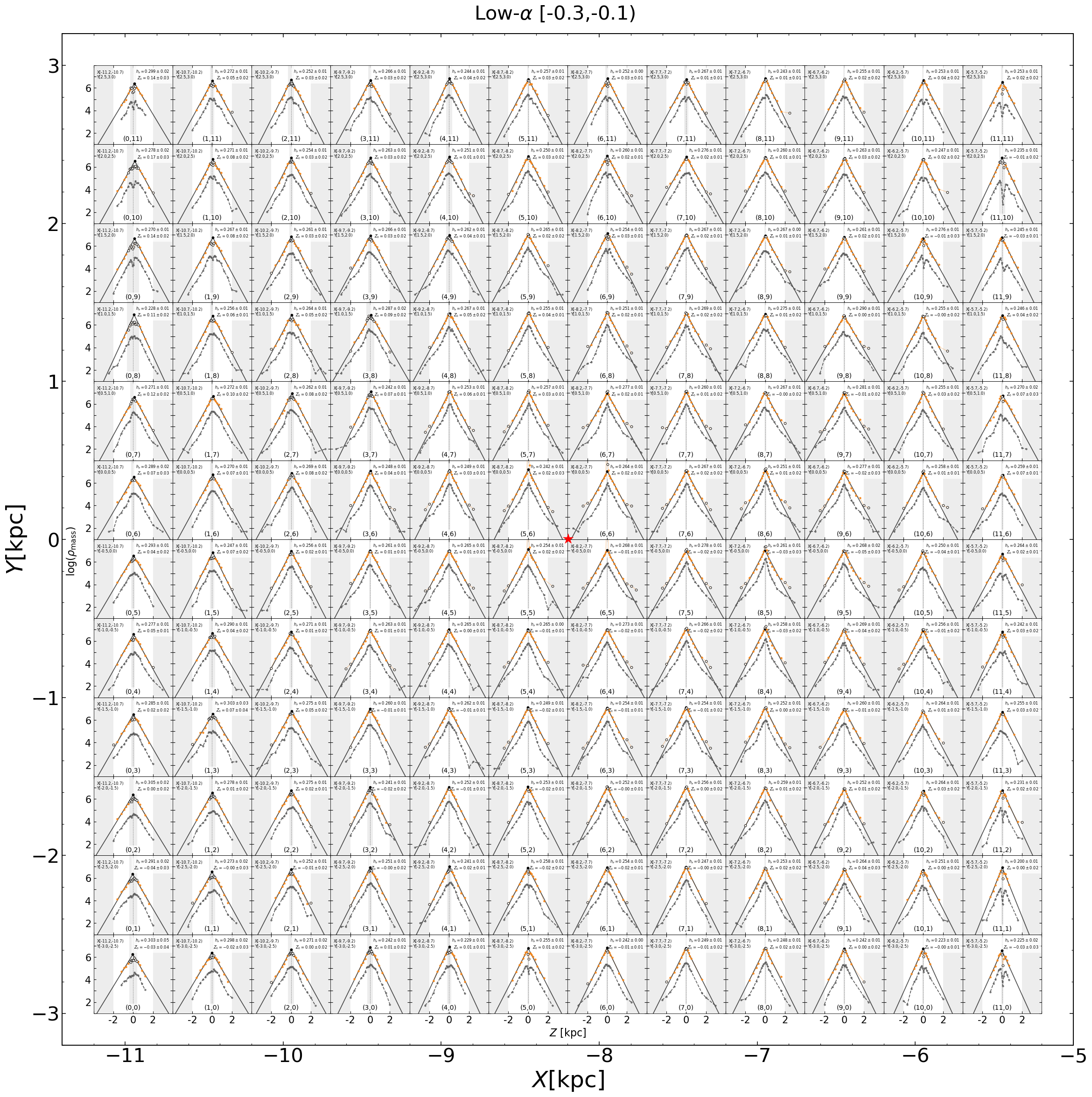}
    \caption{continued.}
    \end{figure}
    
    \begin{figure}[!htbp]
    \ContinuedFloat
    \centering
    \includegraphics[width=\textwidth,height=0.80\textheight,keepaspectratio]{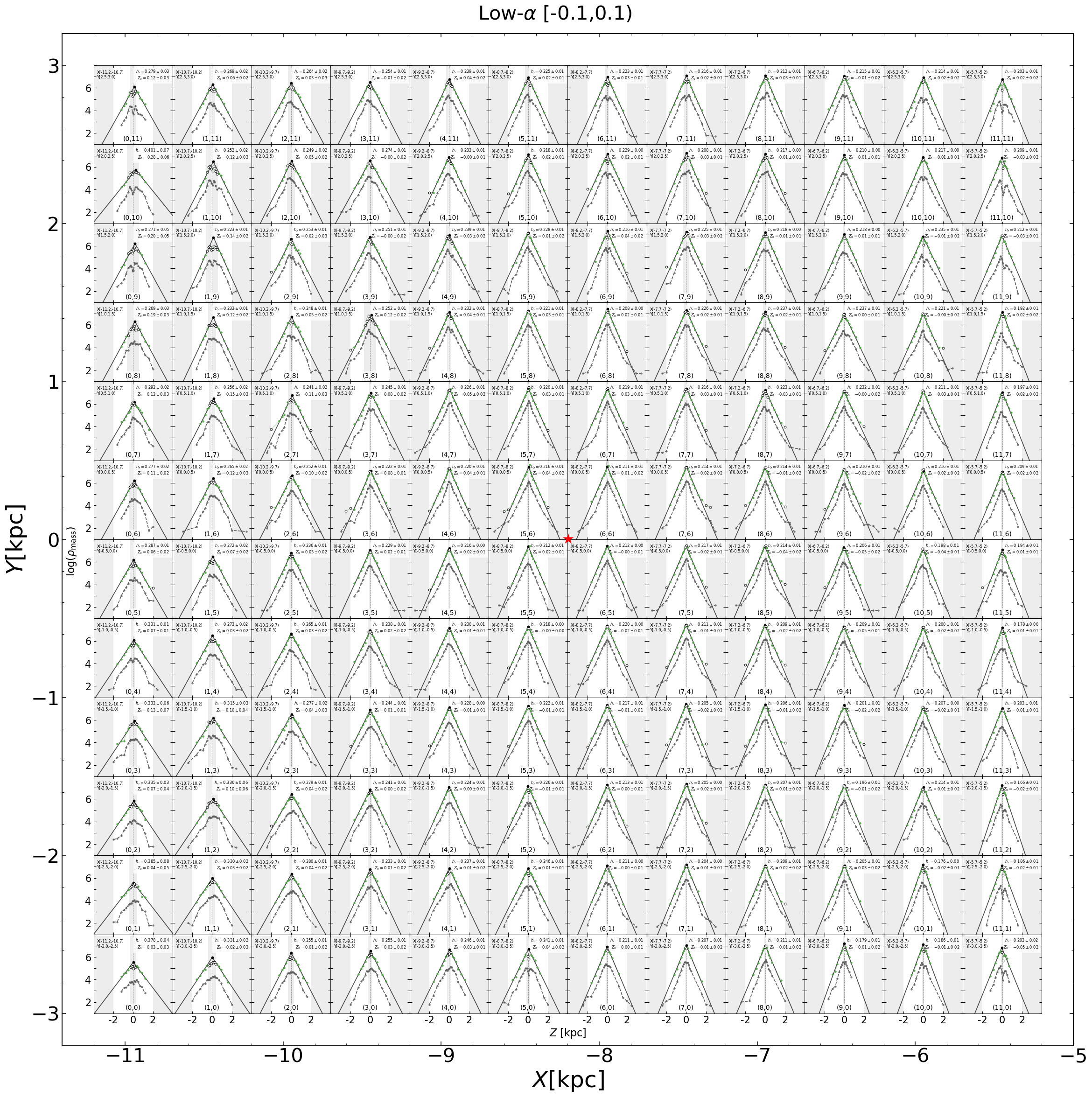}
    \caption{continued.}
    \end{figure}
    
    \begin{figure}[!htbp]
    \ContinuedFloat
    \centering
    \includegraphics[width=\textwidth,height=0.80\textheight,keepaspectratio]{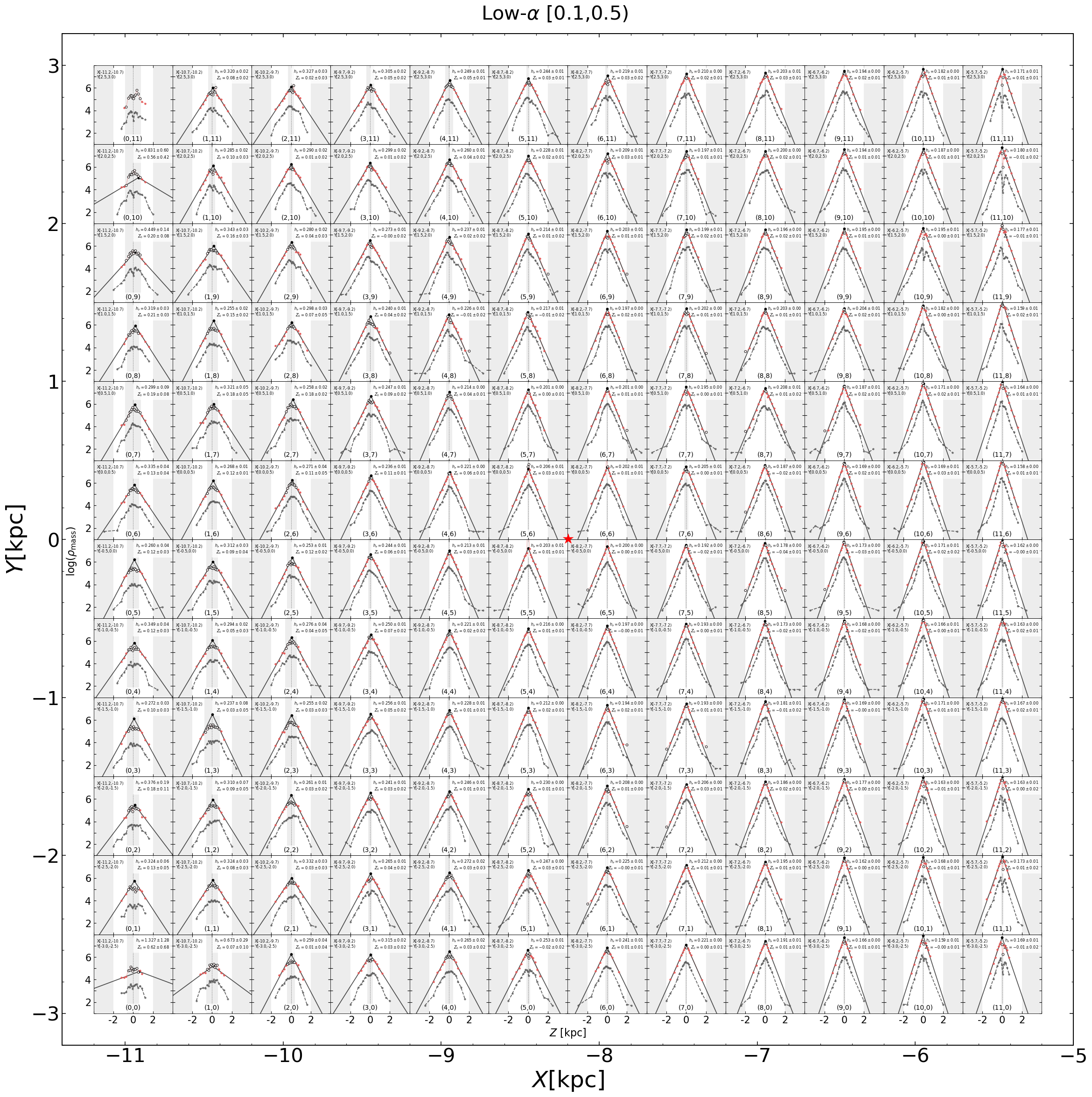}
    \caption{continued.}
    \end{figure}
    \clearpage

\section{Additional Figures}
\label{app:chemical}

\begin{figure}[!htbp]
    \centering
    \includegraphics[width=\textwidth,height=0.85\textheight,keepaspectratio]{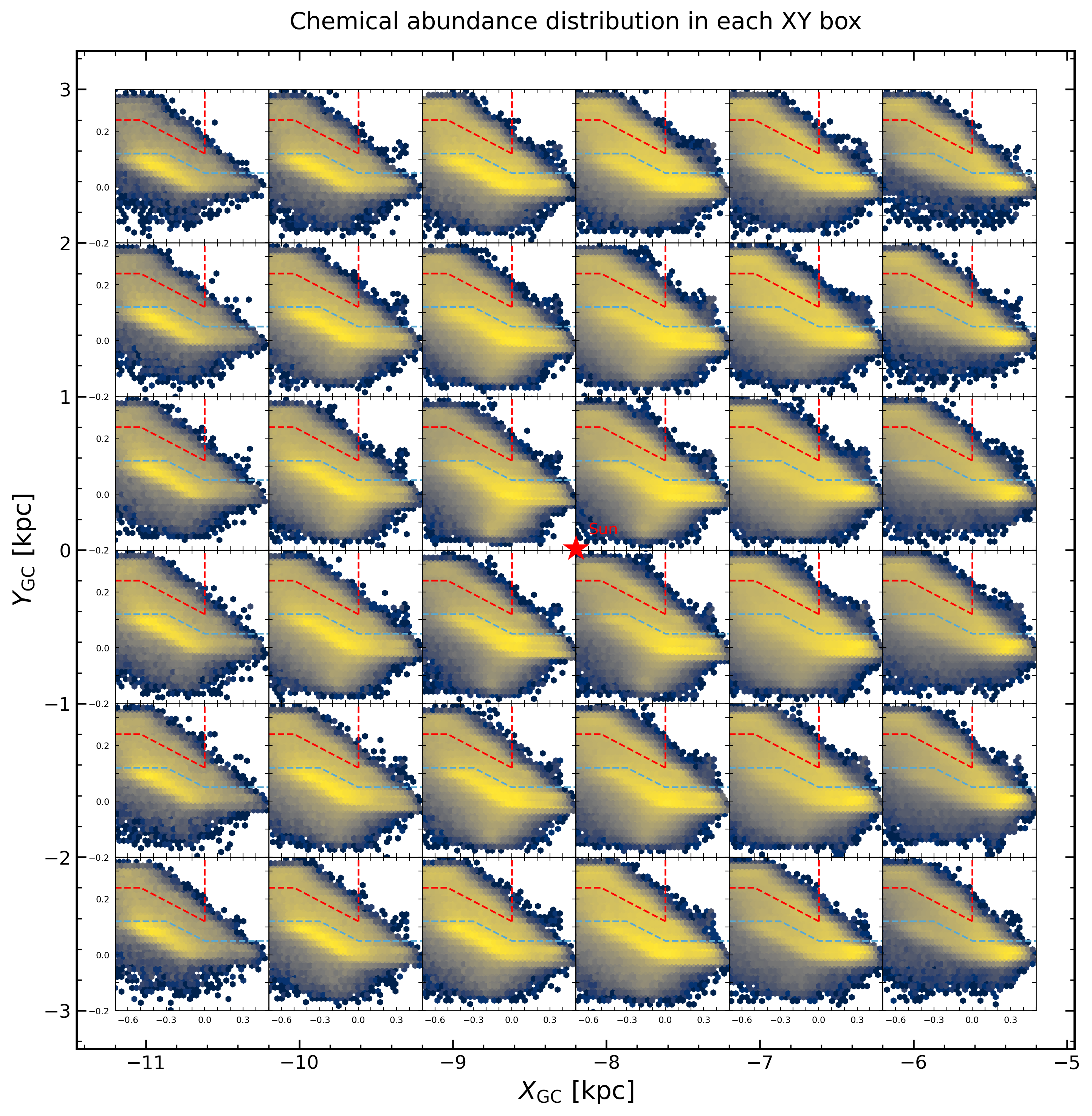}
    \caption{Spatial variation of the $[\alpha/{\rm M}]$--$[{\rm M/H}]$ abundance distribution in the Galactocentric $XY$ plane. Each panel corresponds to a $1\,{\rm kpc}\times1\,{\rm kpc}$ spatial cell. The red and blue dashed lines indicate the adopted high-$\alpha$ and low-$\alpha$ classification boundaries, respectively, as defined by Eqs.~(\ref{eq:alpha_high}) and~(\ref{eq:alpha_low}).}
    \label{fig:chemical_xy}
    \end{figure}

    This section provides additional diagnostic figures used to assess the chemical classification adopted in this work. Figure~\ref{fig:chemical_xy} shows the spatial variation of the $[\alpha/{\rm M}]$--$[{\rm M/H}]$ abundance distribution in the Galactocentric $XY$ plane, which is used to examine possible spatial variations and contamination in the chemical selection of different populations.
    
    \clearpage
    
\end{appendix}
\end{document}